\documentclass[preprint]{aastex}

\newcommand{\HA}{\ensuremath{\mathrm{H}\alpha}}
\newcommand{\HB}{\ensuremath{\mathrm{H}\beta}}
\newcommand{\OIIlam}{[O {\sc ii}]$\lambda3727$}
\newcommand{\OIII}{[O {\sc iii}]$\lambda\lambda$4959,5007}

\newcommand{\OII}{[O {\sc ii}]}

\begin{document}

\title{THE \OIIlam\ LUMINOSITY FUNCTION AND STAR FORMATION RATE AT $z \approx 1.2$ 
IN THE COSMOS 2 SQUARE-DEGREE FIELD AND THE SUBARU DEEP FIELD\altaffilmark{1}}

\author{M. I. Takahashi   \altaffilmark{2},
	    Y. Shioya     \altaffilmark{3},
        Y. Taniguchi  \altaffilmark{3},
        T. Murayama    \altaffilmark{2},
        M. Ajiki        \altaffilmark{2},
        S. S. Sasaki    \altaffilmark{3},
        O. Koizumi	    \altaffilmark{2},
        T. Nagao         \altaffilmark{4, 5},
        N. Z. Scoville      \altaffilmark{6, 7},
        B. Mobasher         \altaffilmark{8},
        H. Aussel           \altaffilmark{7,9},
	 P. Capak          \altaffilmark{6},
	 C. Carilli		  \altaffilmark{10},
	 R. S. Ellis		  \altaffilmark{6},
	 B. Garilli		  \altaffilmark{11},
	 M. Giavalisco     \altaffilmark{8},
	 L. Guzzo          \altaffilmark{12},
	 G. Hasinger       \altaffilmark{13},
	 C. Impey          \altaffilmark{14},
	 M. G. Kitzbichler		\altaffilmark{15},
	 A. Koekemoer		\altaffilmark{8},
	 O. Le F$\grave{{\rm e}}$vre  \altaffilmark{16},
	 S. J. Lilly          \altaffilmark{17},
	 D. Maccagni			 \altaffilmark{11},
	 A. Renzini        \altaffilmark{18},
	 M. Rich            \altaffilmark{19},
	 D. B. Sanders      \altaffilmark{7},
	 E. Schinnerer       \altaffilmark{20},
	 M. Scodeggio        \altaffilmark{11},
	 P. Shopbell       \altaffilmark{6},
	 V. Smolcic		   \altaffilmark{21, 22},
	 S. Tribiano        \altaffilmark{23,24},
        Y. Ideue   \altaffilmark{3},
        S. Mihara \altaffilmark{3}
}
        		
\altaffiltext{1}{Based on data collected at Subaru Telescope, which is operated by 
	             the National Astronomical Observatory of Japan.}
\altaffiltext{2}{Astronomical Institute, Graduate School of Science,
                 Tohoku University, Aramaki, Aoba, Sendai 980-8578, Japan}
\altaffiltext{3}{Physics Department, Graduate School of Science \& Engineering, Ehime University, 
                 Bunkyo-cho, Matsuyama 790-8577, Japan}
\altaffiltext{4}{National Astronomical Observatory of Japan, 
	             Mitaka, Tokyo 181-8588, Japan}
\altaffiltext{5}{INAF --- Osservatorio Astrofisico di Arcetri, 
                 Largo Enrico Fermi 5, 50125 Firenze, Italy}
\altaffiltext{6}{California Institute of Technology, MC 105-24, 1200 East
                 California Boulevard, Pasadena, CA 91125}
\altaffiltext{7}{Institute for Astronomy,  University of Hawaii,
                 2680 Woodlawn Drive, Honolulu, HI 96822}
\altaffiltext{8}{Space Telescope Science Institute, 3700 San Martin Drive, 
                 Baltimore, MD 21218}
\altaffiltext{9}{Service d'Astrophysique, CEA/Saclay, 91191 
                 Gif-sur-Yvette, France}
\altaffiltext{10}{National Radio Astronomy Observatory,
                  P.O. Box 0, Socorro, NM 87801-0387}
\altaffiltext{11}{CNR Instituto di Astrofisica Spaziale e Fisica Cosmica Milano, 
				  Milan, Italy} 
\altaffiltext{12}{Osservatorio Astronomico di Brera, via Brera,
                  Milan, Italy}
\altaffiltext{13}{Max Planck Institut fuer Extraterrestrische
                  Physik, D-85478 Garching, Germany}
\altaffiltext{14}{Steward Observatory, University of Arizona,
                  933 North Cherry Avenue, Tucson, AZ 85721}
\altaffiltext{15}{Max-Planck-Institut f\"ur Astrophysik,
                  D-85748 Garching bei M\"unchen, Germany}                 
\altaffiltext{16}{Laboratoire d'Astrophysique de Marseille,
                  BP 8, Traverse du Siphon, 13376 Marseille Cedex 12, France}
\altaffiltext{17}{Department of Physics, Swiss Federal Institute
                  of Technology (ETH-Zurich), CH-8093 Zurich, Switzerland}
\altaffiltext{18}{European Southern Observatory,
                  Karl-Schwarzschild-Str. 2, D-85748 Garching, Germany}
\altaffiltext{19}{Department of Physics and Astronomy,
                  University of California, Los Angeles, CA 90095}
\altaffiltext{20}{Max Planck Institut f\"ur Astronomie,
                  K\"onigstuhl 17, Heidelberg, D-69117, Germany}
\altaffiltext{21}{Princeton University Observatory, Princeton, NJ 08544}
\altaffiltext{22}{University of Zagreb, Department of Physics, Bijenicka cesta 32, 
                  10000 Zagreb, Croatia}
\altaffiltext{23}{American Museum of Natural History}
\altaffiltext{24}{CUNY Bronx Community College, New York, NY}

\shortauthors{Takahashi et al.}
\shorttitle{\OII\ luminosity function at $z \approx 1.2$}

\begin{abstract}
We have carried out a wide-field imaging survey for \OIIlam\ emitting 
galaxies at $z \approx 1.2$ in the HST COSMOS 2 square degree field using 
the Suprime-Cam on the Subaru Telescope. 
The survey covers a sky area of 6700 arcmin$^2$ in the COSMOS field, 
and a redshift range between 
1.17 and 1.20 ($\Delta z = 0.03$), corresponding to a survey volume of 
5.56$\times 10^5$ Mpc$^3$. 
We obtain a sample of 3176 [O {\sc ii}] emitting galaxies with  observed 
emission-line equivalent widths greater than 26 \AA. 
Since our survey tends to sample brighter \OII\ emitting galaxies, 
we also analyze a sample of fainter \OII\ emitting galaxies found in
the Subaru Deep Field (SDF). 
We find an extinction-corrected [O {\sc ii}] luminosity density of 
$10^{40.35^{+0.08}_{-0.06}}$ ergs s$^{-1}$ Mpc$^{-3}$, 
corresponding to star formation rate density of 
$0.32^{+0.06}_{-0.04}$ $M_{\sun}$ yr$^{-1}$ Mpc$^{-3}$ 
in the COSMOS field at $z \approx 1.2$. 
This is the largest survey for \OII\ emitters beyond $z = 1$ currently available.
\end{abstract}
\keywords{galaxies: distances and redshifts --- galaxies: evolution --- 
          galaxies: luminosity function, mass function}


\section{INTRODUCTION}

Measurements of the galaxy star formation rate density (SFRD) are one of the most
important issues concerning studies of formation and evolution of galaxies. 
Since the pioneering study by Madau et al. (1996), 
a number of observational studies have been made to investigate 
both metal-enrichment history and star formation history in galaxies 
as a function of age or redshift (e.g., Gallego et al. 1995; 
Ellis et al. 1996; Lilly et al. 1996; Hogg et al. 1998; Tresse \& Maddox 1998;
Madau, Pozzetti, \& Dickinson 1998; 
Pettini et al. 1998; Steidel et al. 1999; Barger et al. 2000; Fujita et al. 2003a; 
Giavalisco et al. 2004; Dickinson et al. 2004; Taniguchi et al. 2005; Bouwens \&
Illingworth 2006).  
A current picture is that the SFRD steeply increases for first 900 Myrs
(from $z \sim 30$ to $z \sim 6$), peaks at $z \sim 3$, and then decreases strongly 
to the present day (e.g., Bouwens \& Illingworth 2006).  
It is important to investigate what happened in first 900 Myrs in early universe
(Taniguchi et al. 2005; Kashikawa et al. 2006; Bouwens \& Illingworth 2006
and references therein). And, it is also important to investigate what happened from
$z \sim 3$ to $\sim 1$  because galaxies as well as large-scale structures
could have rapidly evolved in this era.  

Star formation activity in galaxies could be linked to galaxy environments because 
interaction and merging events can trigger intense star formation. Therefore,
it is necessary to investigate co-evolution between galaxies and large-scale 
structures (e.g., Scoville et al. 2007). 
Recently, large-scale structures with a scale of several tens to 100 Mpc
have been reported for samples of Ly$\alpha$ emitters or Lyman break galaxies (LBG);
i.e. star-forming galaxies, at $3 < z < 6$ (e.g., Steidel et al. 1998; 
Shimasaku et al. 2003; Ouchi et al. 2005a; Kashikawa et al. 2006). 
At $z \sim 1$, such large-scale structure of galaxies with a scale over 10 Mpc 
have also been reported (Tanaka et al. 2001; Nakata et al. 2005; Gal, Lubin \& 
Squires 2005). 
These observational results appear to be consistent with $N$-body simulations which 
suggest that clusters of galaxies at $z \sim 1$ are still in the formation 
process (e.g., Moore et al. 1998).
Therefore, more detailed studies on clustering of galaxies at $z > 1$ are
crucially important to understand the evolution of large-scale structures as well 
as that of galaxies themselves.

It is, however, difficult to obtain a large sample of galaxies beyond $z = 1$.
Although optical broad-band imaging could detect a large number of LBGs,
the accuracy of photometric redshifts is not high enough to obtain a sample of 
galaxies at a concerned redshift interval. Spectroscopic surveys are useful
to obtain such a sample. However, since most galaxies beyond $z = 1$ are 
faint, we need a lot of observing time to carry out such a program. 
Another method is to carry out an optical survey with a narrowband filter to isolate
strong emission-line galaxies beyond $z = 1$.
However, beyond $z \sim 0.5$, the H$\alpha$ emission line, which is a good tracer 
of star formation rate  (e.g., Gallego et al. 1996; Hammer et al. 1997; Jansen et al. 
2001; Teplitz et al. 2003; Hopkins 2004 and references therein), is no longer 
detectable in the visible window. Therefore, the \OIIlam\ emission line
provides an estimator of the star-formation activity in galaxies.
Although this line is considered to be a less reliable tracer of star formation rate 
because its line intensity is more affected by dust obscuration (Kennicutt 1998), 
careful comparisons among SFRs derived with various observables (e.g., H$\alpha$,
[O {\sc ii}], UV, and so on) are consistent within a factor of three (Cram et al.
1998; Hopkins 2004).
Therefore, this line has the clear advantage of being observable in the visible
bands over the interesting redshift range of $0.5 < z < 1.6$,
where the SFRD is thought to be changing rapidly. 

In this paper, we present results of our narrowband imaging survey for 
[O {\sc ii}] emitters at $z \approx 1.2$ in the Cosmic Evolution Survey 
(COSMOS; Scoville et al. 2007) 2 square degree field.
This survey was carried out during the course of Subaru imaging surveys
of the COSMOS field (Taniguchi et al. 2007; see also Murayama et al. 2007)
by using  the narrowband filter {\it NB}816 [$\lambda_{c}=8150$ \AA~ and 
$\Delta\lambda$(FWHM) = 120 \AA] available for the prime-focus, wide-field camera, 
Suprime-Cam on the 8.2m Subaru Telescope (see Section 2.1). 
Since the Suprime-Cam provides a very wide field of view; $34^\prime \times 27^\prime$, 
this is suitable for wide-field optical imaging surveys. 
Through the large area covered by the COSMOS, we can obtain the largest sample of \OII\ 
emitting galaxies beyond $z = 1$ by performing narrowband imaging of this field,
targeting galaxies at $z \approx 1.2$. Such a large 
sample can improve the precision of our estimates of the SFRD at $z > 1$. 

However, the COSMOS can sample only brighter \OII\ emitting galaxies 
because of relatively shallower survey depths. 
In order to obtain an accurate luminosity function of [O {\sc ii}] emitters, 
it is also necessary to probe faint [O {\sc ii}] emitters. 
For this purpose, 
we analyze a sample of fainter \OII\ emitters taken from 
the Subaru Deep Field (SDF; Kashikawa et al. 2004) where the same narrowband
filter ({\it NB}816) was used (see Section 2.2). 
Although the SDF covered sky area is $\sim 1 / 8$ of the COSMOS field, 
the much deeper survey depth of the SDF makes it possible to explore faint objects 
and determine the faint end slope of the luminosity function of [O {\sc ii}] emitting 
galaxies. 

Throughout this paper, magnitudes are given in the AB system. 
We adopt a flat universe with the following cosmological parameters;
$\Omega_{\rm matter} = 0.3$, 
$\Omega_{\Lambda} = 0.7$, and $H_0 = 70 \; {\rm km \; s^{-1} \; Mpc^{-1}}$. 

\section{PHOTOMETRIC CATALOG}

\subsection{The HST COSMOS Field}

The Cosmic Evolution Survey (COSMOS) is a 2 square degree imaging survey centered 
at $\alpha$(J2000)$=10^{\rm h}00^{\rm m}28^{\rm s}.6$ and 
$\delta$(J2000)$=+02^{\circ}12^\prime 21^{\prime \prime}.0$ (Scoville et al. 2007). 
Our optical narrow-band imaging observations of the HST COSMOS field have been made with 
the Suprime-Cam (Miyazaki et al. 2002) on the 8.2 m Subaru Telescope (Kaifu et al. 2000; 
Iye et al. 2004) at Mauna Kea Observatories. 

In this analysis, we use the COSMOS official photometric redshift catalog which 
includes objects whose total $i$ magnitudes ($i^\prime$ or $i^*$) are brighter than 25 mag. 
The catalog presents $3^{\prime \prime}$ diameter aperture magnitude of 
Subaru/Suprime-Cam $B$, $V$, $r^\prime$, $i^\prime$, $z^\prime$, and {\it NB}816
\footnote{Our SDSS broad-band filters are designated as 
$r^+$, $i^+$, and $z^+$ in Capak et al. (2007) to distinguish from the original 
SDSS filters. Also, our $B$ and $V$ filters are designated as $B_J$ and $V_J$ in 
Capak et al. (2007) where $J$ means Johnson and Cousins filter system used in 
Landolt (1992).}. 
Our {\it NB}816 imaging data of the COSMOS field are also used to search both 
for Ly$\alpha$ emitters at $z \approx 5.7$ (Murayama et al. 2007) and for \HA\ emitters 
at $z \approx 0.24$ (Shioya et al. 2007). 
Details of the Suprime-Cam observations are given in Taniguchi et al. (2007). 

Because the accuracy of standard star calibration ($\pm 0.05$ magnitude) is too 
large for obtaining accurate photometric redshifts, 
Capak et al. (2007) re-calibrated the photometric zero-points for photometric redshifts 
using the SEDs of galaxies with spectroscopic redshifts. 
Following the recommendation of Capak et al. (2007), we apply the zero-point correction 
to the photometric data in the official catalog. 
The offset values are 0.189, 0.04, $-$0.040, $-$0.020,$-$0.005, 0.054 and $-$0.072 for 
$B$, $V$, $r^\prime$, $i^\prime$, $i^*$, $z^\prime$, and $NB816$, respectively. 
The zero-point corrected limiting magnitudes are 
$B = 27.4$, $V = 26.5$, $r^\prime = 26.6$, $i^\prime = 26.1$, $z^\prime = 25.4$ 
and $NB816 = 25.6$ for a $3 \sigma$ detection on a $3^{\prime \prime}$ diameter aperture. 
The catalog also includes $3^{\prime \prime}$ diameter $i^*$ band aperture 
magnitude from CFHT. 
We use the CFHT $i^*$ magnitude for bright galaxies with $i^\prime < 21$ mag 
because they appear to be less affected by the saturation effects. 
All magnitudes are corrected for the Galactic extinction; 
$\bar{E}(B - V) = 0.0195 \pm 0.006$ (Capak et al. 2007). 
Details of the COSMOS official photometric redshift catalog
is presented in Mobasher et al. (2006).

\subsection{The Subaru Deep Field}

We used a catalog obtained by the Subaru Deep Field (SDF) project (Kashikawa et al. 2004). 
The SDF project is the deepest optical imaging survey using the Suprime-Cam 
on the Subaru Telescope. The SDF is located near the North Galactic Pole, 
being centered at $\alpha$(J2000)$=13^{\rm h} 24^{\rm m} 38^{\rm s}.9$ and 
$\delta$(J2000)$=+27^\circ29^\prime 25^{\prime\prime}.9$. 
The SDF project official photometric catalog is obtained from the SDF site 
(http://step.mtk.nao.ac.jp/sdf/project/). 
In this work, we used $B$, $V$, $R_{c}$, $i^\prime$, $z^\prime$, and {\it NB}816 data
given in the $i^\prime$ selected catalog with $3^{\prime \prime}$ diameter aperture photometry.
Note that we correct {\it NB}816 magnitudes in the SDF photometric catalog
because there is a small photometric offset ($\approx 0.08$ mag). 
The limiting magnitudes for a $3\sigma$ detection on a $3^{\prime \prime}$ 
diameter aperture are as follows; $B = 27.6$, $V = 27.0$, $R_{c} = 27.2$, $i^\prime = 27.0$, 
$z^\prime = 26.2$, and ${\it NB}816 = 26.1$ mag. The PSF size in this catalog is 0.98$\arcsec$.
All magnitudes are corrected for the Galactic extinction;
$E(B - V) = 0.01678 \pm 0.003$ (Schlegel et al. 1998). 
This complements our COSMOS [O {\sc ii}] survey by extending it to fainter flux limits.

Table \ref{oii:tab:sdfcosmos} summarizes the covered sky area, survey volume and 
photometric property in the COSMOS field and the SDF. 
Since the central wavelength of {\it NB}816 filter corresponds to a redshift of 1.2 
for \OIIlam\ emission, together with the broadband filter data, 
we can carry out a search for [O {\sc ii}] emitters at $z \approx 1.2$. 

\section{RESULTS}

\subsection{Selection of {\it NB}816-Excess Objects}

In order to select {\it NB}816-excess objects efficiently, 
it is desirable to have a wavelength-matched continuum image as off-band data. 
Since the central wavelength of the {\it NB}816 filter is different either from 
those of $i$ ($i^\prime$ or $i^*$) and $z^\prime$ filters, we make a 
wavelength-matched continuum, ``$iz$ continuum'', using the following 
linear combination; $f_{iz} = 0.57 f_{i}+0.43 f_{z^\prime}$ 
where $f_{i}$ and $f_{z^\prime}$ are the $i$ ($i^\prime$ or $i^*$) and $z^\prime$ 
flux densities, respectively. Its 3 $\sigma$ limiting magnitude is $iz \simeq 26.0$ 
mag in a $3^{\prime \prime}$ diameter aperture in the COSMOS field, 
while $26.9$ mag in the SDF. Note that we use the CFHT $i^*$ magnitude 
for the bright galaxies with $i^\prime < 21$ mag in the COSMOS field
since their $i^\prime$ magnitudes suffer from saturation effects. 
This enables us to more precisely sample the continuum at the same 
effective wavelength as that of the {\it NB}816 filter. 

\begin{figure}
\begin{center}
\epsscale{1.3}
\plottwo{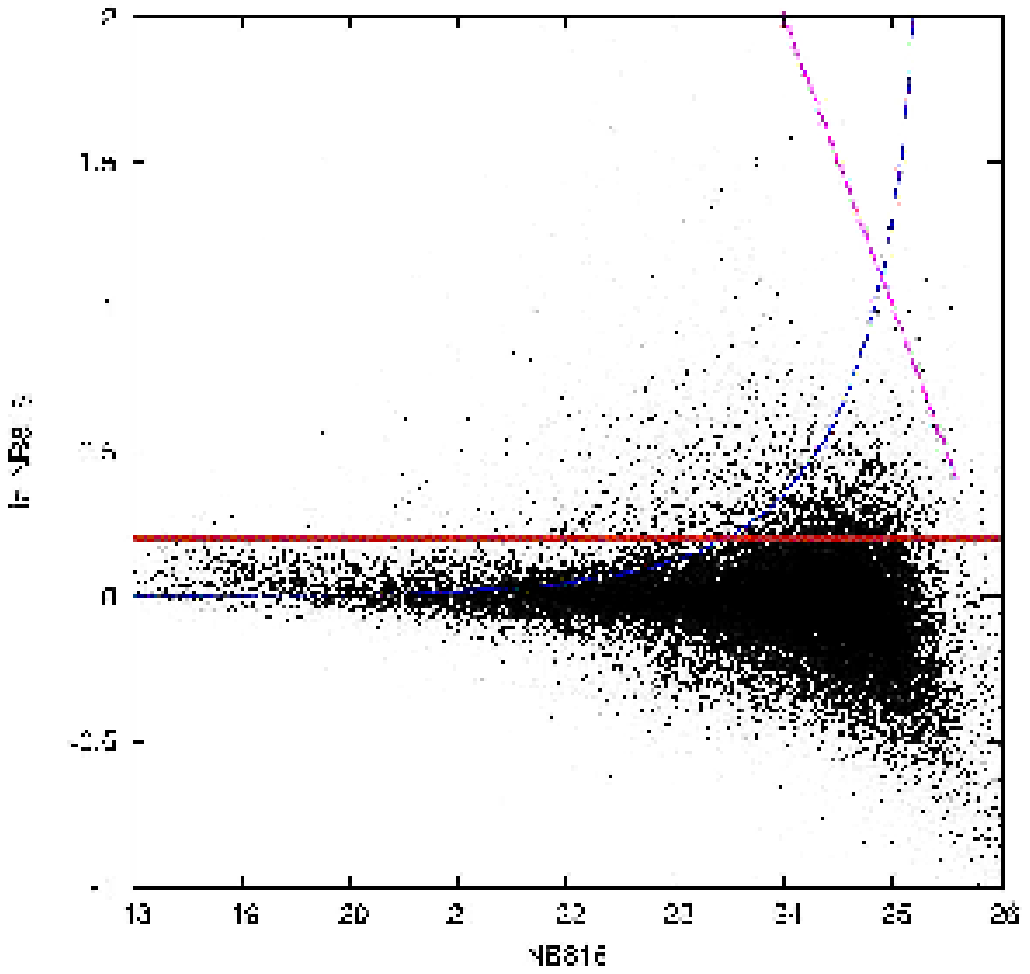}{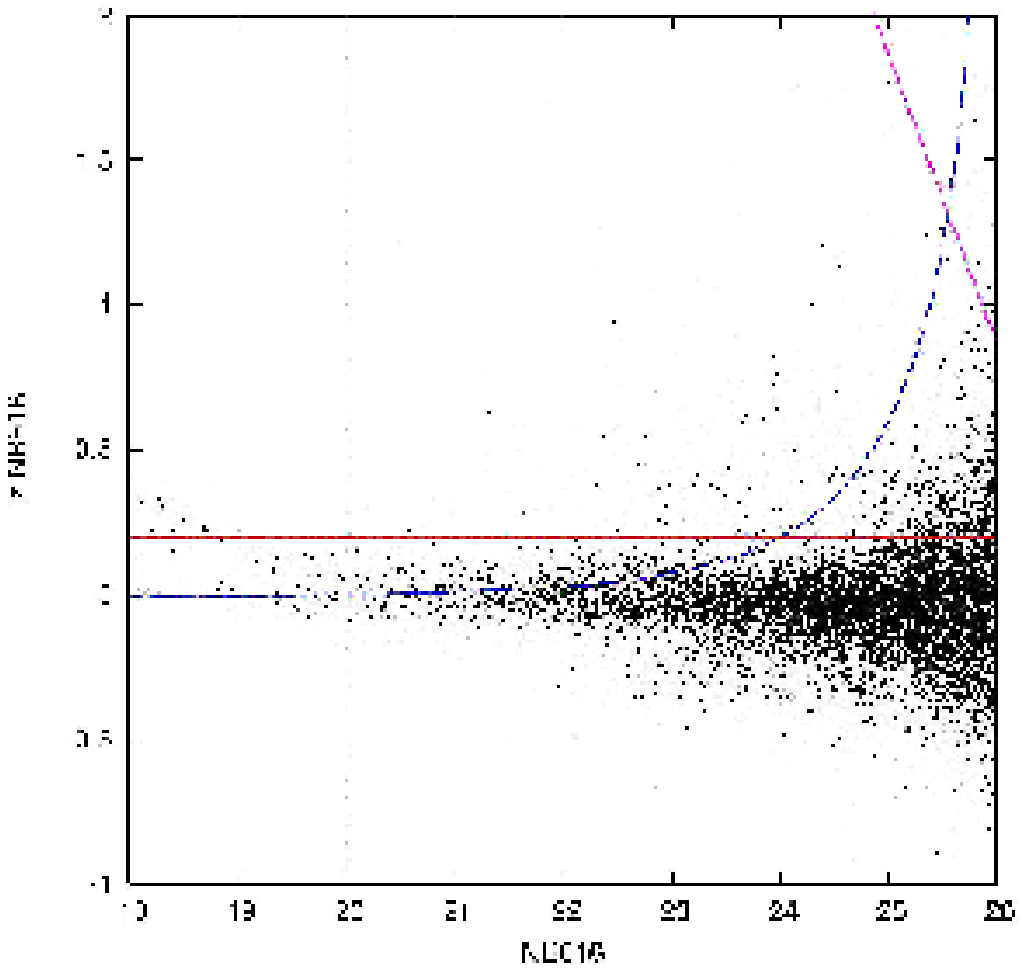}
\caption{Diagrams between $iz - \mathit{NB816}$ and {\it NB}816  
for all objects in the $i^\prime$-selected catalogue in the COSMOS field 
($Top$) and the SDF ($Bottom$). 
The horizontal solid red line corresponds to $iz - \mathit{NB816} = 0.2$. 
The blue line shows the distribution of $3\sigma$ error. 
The pink line shows the limiting magnitude of $iz$.
The dotted line shows ${\it NB}816 = 20$ mag.  
\label{Ha:iz-NBvsNB}}
\end{center}
\end{figure}

Taking both the scatter in the $iz - \mathit{NB816}$ 
color and our survey depth into account, candidate line-emitting objects are 
selected with the following criteria.

\begin{equation}
iz - {\mathit NB816} \geq 0.2
\end{equation}
and
\begin{equation} 
iz - {\mathit NB816} > 3 \sigma_{iz-\mathit{NB}816},
\end{equation}
where  $iz - {\mathit NB816} = 0.2$ corresponds to $EW_{\rm obs} \approx 26$ \AA. 
We will justify these adopted criteria in Section 3.3. 
We compute the $3\sigma$ of the color as 
\begin{equation}
3\sigma_{iz-\mathit{NB}816}=-2.5\log
(1-\sqrt{(f_{3\sigma_{\mathit{NB}816}})^2+(f_{3\sigma_{iz}})^2}/f_{\mathit{NB}816}).
\end{equation} 
These two criteria, (1) and (2), are shown in Figure \ref{Ha:iz-NBvsNB} by the red 
and blue solid lines, respectively.
 
In the COSMOS field, taking account of the homogeneity of the noise level, 
we select galaxies in the following region: 
$149^\circ \! \! .40917 < {\rm RA} < 150^\circ \! \! .82680$ 
and $1^\circ \! \! .49056 < {\rm Dec} < 2^\circ \! \! .90705$. 
The effective survey area is 6700 arcmin$^2$. 
We find 5824 sources in the COSMOS field that satisfy the above criteria.

In the case of the SDF, objects with ${\it NB}816 < 20$ mag appear to be saturated.
In order to avoid such saturated bright objects, we have also imposed the
criterion ${\it NB}816 > 20$ mag to the SDF sample. 
We then find 602 sources that satisfy the above criteria in the SDF. 

Note that the above {\it NB}816-excess objects found in both the COSMOS field and 
the SDF are brighter than the limiting magnitude in all bands.

\subsection{Selection of {\it NB}816-Excess Objects at $z \approx 1.2$}

A narrowband survey of emission-line galaxies can potentially 
detect galaxies with different emission lines at different redshifts. 
Strong emission lines that we would expect to detect are H$\alpha$, H$\beta$, 
\OIII\, and \OIIlam\ (Tresse et al. 1999; Kennicutt 1992b). 
In Table \ref{Ha:tab:cover} we show redshift coverages of those emission lines for 
the {\it NB}816 filter.

\begin{figure}
\begin{center}
\epsscale{1.3}
\plottwo{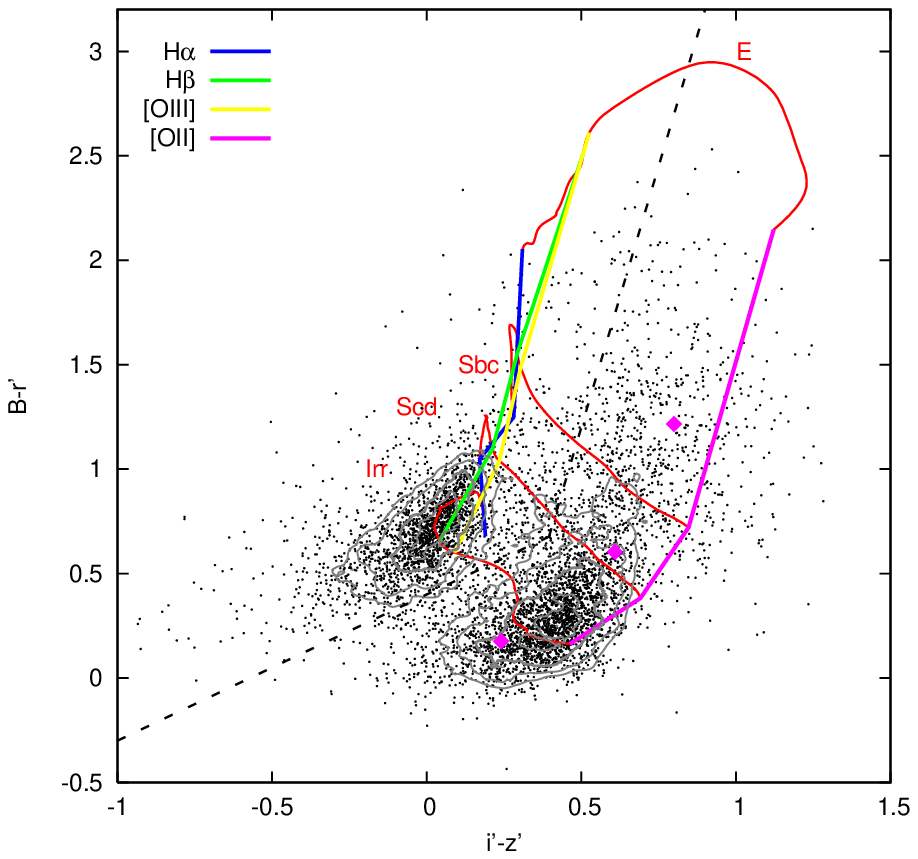}{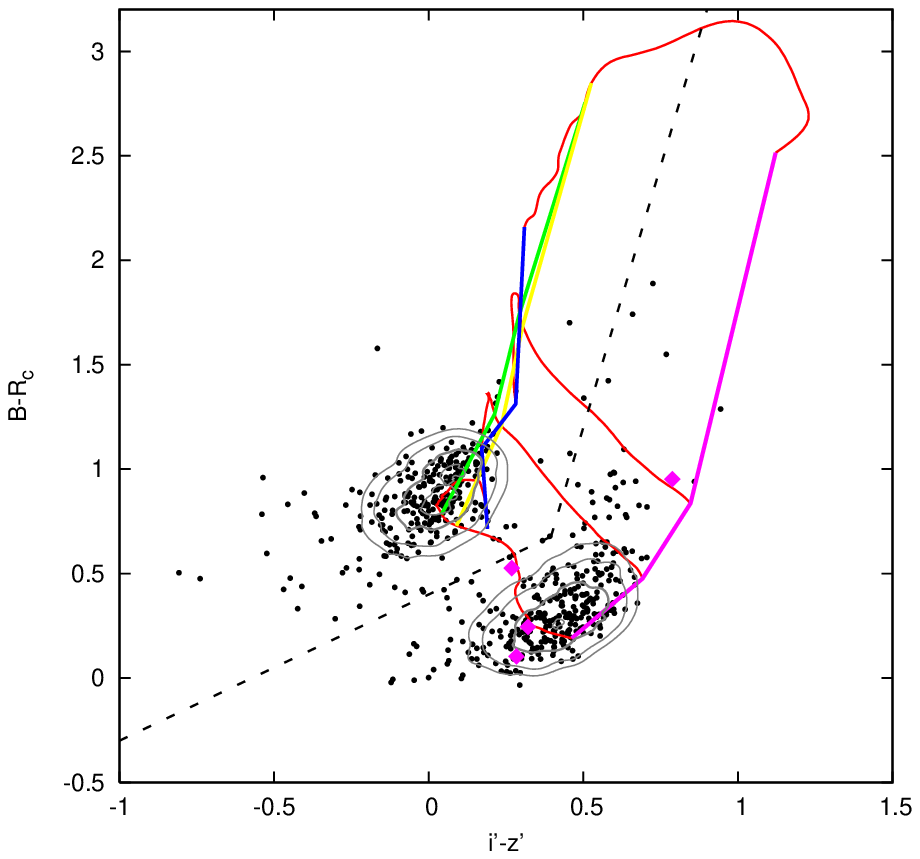}
\caption{Diagrams between $B - r^\prime$ vs. $i^\prime - z^\prime$ for 
the 5824 emitter sources in the COSMOS field ($Top$) and $B - R_{c}$ vs. $i^\prime - z^\prime$ for 
the 602 emitter sources in the SDF ($Bottom$). 
Color loci of model galaxies (E, Sbc, Scd, Irr) from $z = 0$ to $z = 1.2$ are shown with 
red lines. Colors of $z = 0.24$, $0.64$, $0.68$, and $1.18$ 
(for H$\alpha$, [O {\sc iii}], H$\beta$ and [O {\sc ii}] emitters, respectively) 
are shown with blue, yellow, green, and pink lines, respectively. 
The contour levels shown with gray lines corresponds to 2$\mu$, $\mu$, $\mu$/2 and $\mu$/3, 
where $\mu$ is the mean surface density on each diagram. 
Pink points are \OII\ emitters at $z \approx 1.2$ confirmed by spectroscopy. 
We select the sources under the dotted line in each panel as [O {\sc ii}] emitters.
\label{Ha:BVrcolor}}
\end{center}
\end{figure}

\begin{figure}
\begin{center}
\epsscale{1.3}
\plottwo{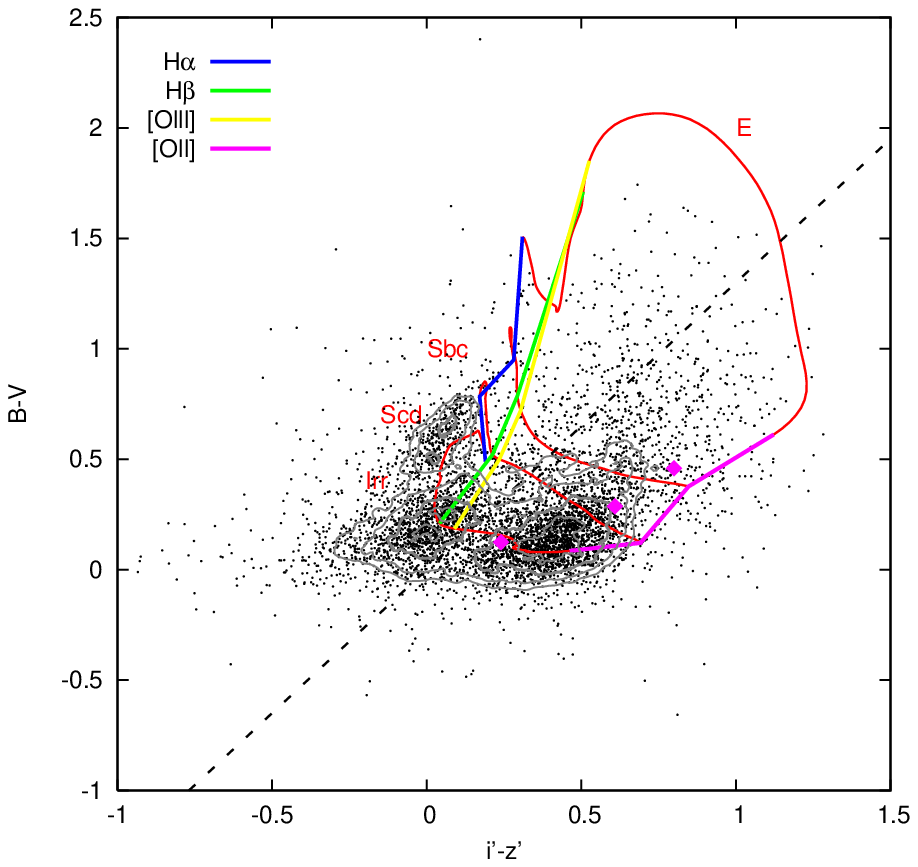}{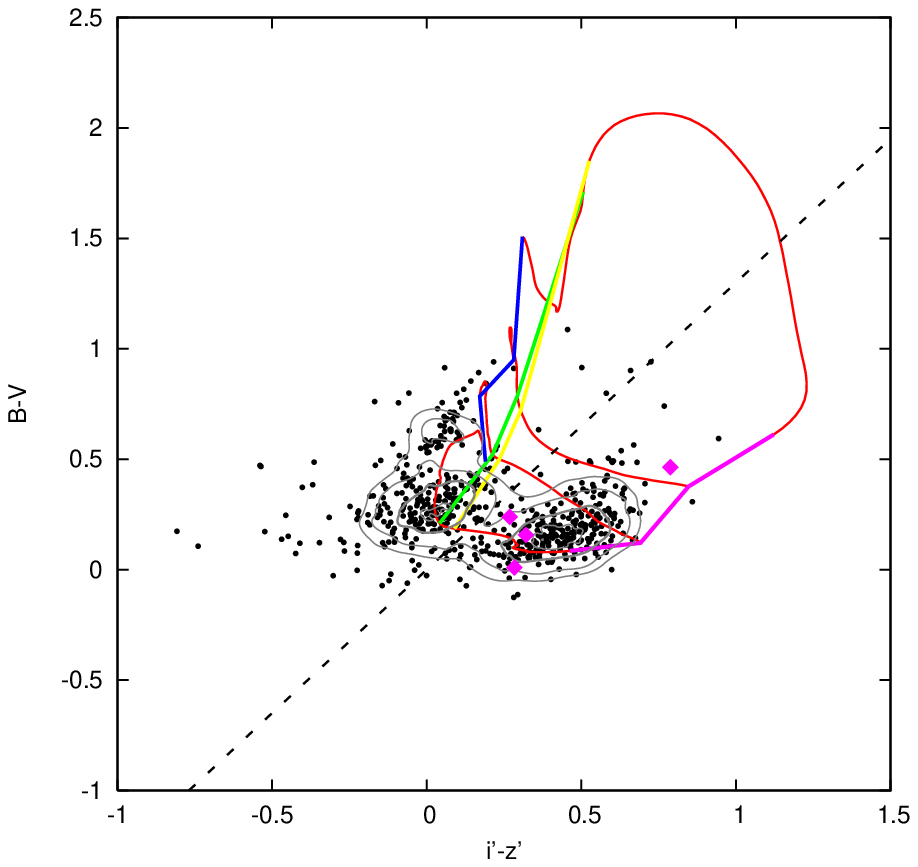}
\caption{Diagrams between $B - V$ vs. $i^\prime - z^\prime$ for 
the 5824 emitter sources in the COSMOS field ($Top$) and the 602 emitter sources in the SDF ($Bottom$). 
The model lines, points and contour levels are same as Figure \ref{Ha:BVrcolor}.
We select the sources under the dotted line in each panel as [O {\sc ii}] emitters.  
\label{Ha:Brizcolor}}
\end{center}
\end{figure}

In order to distinguish \OIIlam\ emitters at $z \approx 1.2$ from 
other emission-line objects at different redshifts, we investigate their 
broad-band color properties. In the upper panels of Figures \ref{Ha:BVrcolor} and 
\ref{Ha:Brizcolor}, we show the $B - r^\prime$ vs. $i^\prime - z^\prime$ and 
$B - V$ vs. $i^\prime - z^\prime$ color-color diagram 
for the 5824 sources in the COSMOS field together with the loci of model galaxies 
taken from Colman, Wu, \& Weedman (1980). 
We can clearly see some clumps in the diagrams. 
We also calculated local surface density on the diagrams 
shown by the contours. The contour levels correspond to 
2$\mu$, $\mu$, $\mu$/2 and $\mu$/3, where $\mu$ corresponds to 
the mean surface density on each color-color diagram. 
These contours show prominent sequences close to the model predictions for 
\OII\ emitters, with distinct, H$\alpha$ or H$\beta$ or \OIII\ emitter 
in the diagrams.
 
Taking account of the model lines and the surface density contours 
on the diagram, we find that \OII\ emitters at $z \approx 1.2$ can be selected by
adopting the following criteria on the $B - r^\prime$ vs. $i^\prime - z^\prime$ color-color diagram; 
\begin{equation}
B - r^\prime < 5(i^\prime - z^\prime) - 1.3, 
\end{equation} 
\noindent or,
\begin{equation}
B - r^\prime < 0.7(i^\prime - z^\prime) + 0.4
\end{equation}
and on the $B - V$ vs. $i^\prime - z^\prime$ color-color diagram; 
\begin{equation}
B - V < 1.3(i^\prime - z^\prime).
\end{equation}  
The candidates which satisfy both of the criteria are selected to be \OII\ emitters at $z \approx 1.2$. 
These criteria give us a sample of 3178 \OII\ emitting galaxy candidates in the COSMOS field. 
We can distinguish [O {\sc ii}] emitters from H$\alpha$ or H$\beta$ or [O {\sc iii}] emitters 
using these criteria. 
Since the two brightest [O {\sc ii}] emitters seem to be AGNs, we do not include them 
in our final samples. Therefore, 
we finally obtain a sample of 3176 \OII\ emitting galaxy candidates in the COSMOS field. 
Note that three \OII\ emitters at $z \approx 1.2$ confirmed by spectroscopy 
(pink points in the upper panels of Figures 2 and 3; see section 3.3) are satisfied 
with the three criteria.

Next, we distinguish \OII\ emitters at $z \approx 1.2$ from 
emission-line objects at other redshifts in the SDF.
In the lower panels of Figures \ref{Ha:BVrcolor} and \ref{Ha:Brizcolor}, 
we show the $B - R_c$ vs. $i^\prime - z^\prime$ and 
$B - V$ vs. $i^\prime - z^\prime$ color-color diagram 
for the 602 sources in the SDF. The contour levels are the same as above. 
Taking account of the model lines and contours of distribution 
on the diagram, 
we find that \OII\ emitters at $z \approx 1.2$ can be selected by
adopting the following criteria on the $B - R_c$ vs. $i^\prime - z^\prime$ color-color diagram; 
\begin{equation}
B - R_{c} < 5(i^\prime - z^\prime) - 1.3, 
\end{equation}
\noindent or,
\begin{equation}
B - R_{c} < 0.7(i^\prime - z^\prime) + 0.4,
\end{equation}
and on the $B - V$ vs. $i^\prime - z^\prime$ color-color diagram;
\begin{equation}
B - V < 1.3(i^\prime - z^\prime). 
\end{equation}
The candidates which satisfy both of the criteria are selected to be \OII\ emitters at $z \approx 1.2$.
These criteria give us a sample of 295 \OII\ emitting galaxy candidates in the SDF.
After eye inspection of the ${\it NB}816$ images, we rejected one object, which appears to be noise.
Therefore, we obtain a sample of 294 \OII\ emitting galaxy candidates in the SDF. 
Note that four \OII\ emitters at $z \sim 1.2$ confirmed by spectroscopy 
(pink points in the lower panels of Figures 2 and 3; see section 3.3) are satisfied 
with the three criteria. 

In this way, we obtain our final \OII\ emitting galaxy candidates 
in both COSMOS and SDF. To our knowledge, this is the largest survey 
for \OII\ emitters beyond $z = 1$ currently available. 

\begin{figure}
\begin{center}
\epsscale{0.8}
\plotone{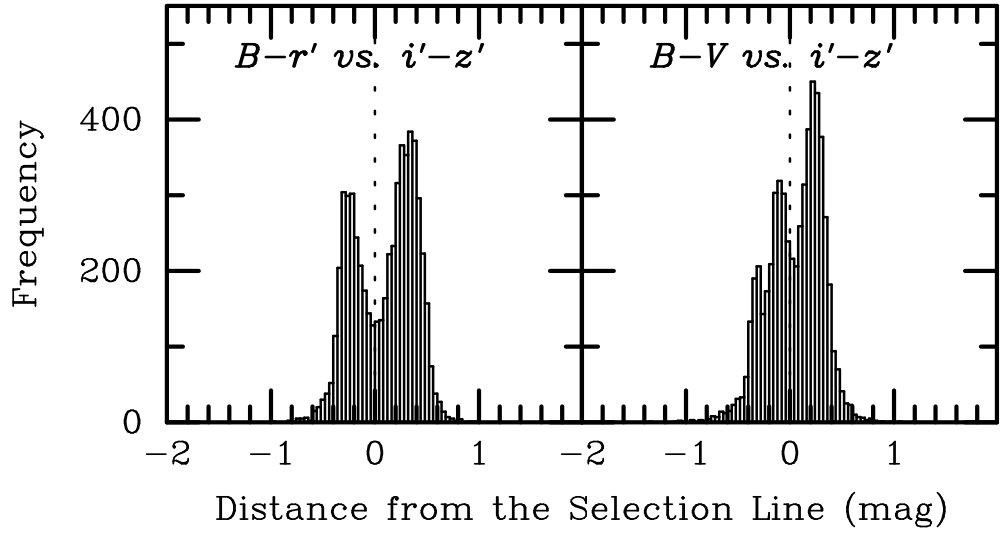}
\caption{The number distributions of emitter candidates on the 
$B - r^\prime$ vs. $i^\prime - z^\prime$ ($left$) and 
$B - V$ vs. $i^\prime - z^\prime$ ($right$) color-color diagrams as a function of distance 
from the adopted selection lines in the COSMOS field. 
The adopted selection lines are shown with dotted lines. The emitter candidates which do not satisfy 
the selection criteria are shown with negative distance, while those satisfy the criteria are shown 
with positive distance. 
\label{dist-col}}
\end{center}
\end{figure}

In order to investigate how our selection criteria (4) - (6) on the color-color 
diagrams affect completeness of the \OII\ emitters and contamination of the 
other emitters, 
we examined the number distribution of emitter candidates as a function of distance 
from the adopted selection line on each color-color diagram in the COSMOS field. 
The three distinct peaks of H$\beta$ + [O {\sc iii}], H$\alpha$, and \OII\ emitters can be 
seen in the distribution of $B - V$ vs. $i^\prime - z^\prime$ color-color diagram 
shown in the right panel of Figure \ref{dist-col} while prominent two peaks 
of \OII\ emitters and that of other emitters can be seen for the 
$B - r^\prime$ vs. $i^\prime - z^\prime$ color-color diagram in the left panel of 
Figure \ref{dist-col}. 
We fit the distributions with the Gaussian and estimate the amount of contamination by 
H$\alpha$ or H$\beta$ + [O {\sc iii}] emitters and incompleteness of the \OII\ emitters. 
Then we find that the amount of contamination and incompleteness are 
$\sim$2\% and $\sim$3\%, respectively. 
Thus, using the two color-color diagrams, our procedures presented 
here can be basically free from contaminations from H$\alpha$ or H$\beta$ or [O {\sc iii}] 
emitters and allows us to select reliable \OII\ emitter samples at $z \approx 1.2$. 
 
\subsection{Optical Spectroscopy}

Our pilot spectroscopic survey for objects in the COSMOS field 
is now progressing (z-COSMOS; Lilly et al. 2006). 
The first results from z-COSMOS have already confirmed that the three galaxies 
(ID 948253, ID 1297800, and ID 1690252)
are star-forming galaxies at $1.17 < z < 1.20$ (see Appendix). 

If we adopted the {\it NB}-excess criterion of $iz - {\mathit NB816} \geq 0.1$ 
and the same color-color selection as we have described in Section 3.2, 
our photometric sample would contain 8 galaxies 
which were spectroscopically observed. 
The four galaxies among them are star-forming galaxies at $1.17 < z < 1.20$. 
However, the remaining four galaxies are all at $z \approx 1.02$ where 
the \OII\ emission line is out of the bandpass of the {\it NB}816 filter. 
Since we find no H$\alpha$ or H$\beta$ or [O {\sc iii}] emitters in our 
spectroscopic sample, our color-color selection criteria are 
reliable discriminators between \OII\ emitters at $z \sim 1.2$ and other emission line objects. 

\begin{figure}
\epsscale{0.6}
\plotone{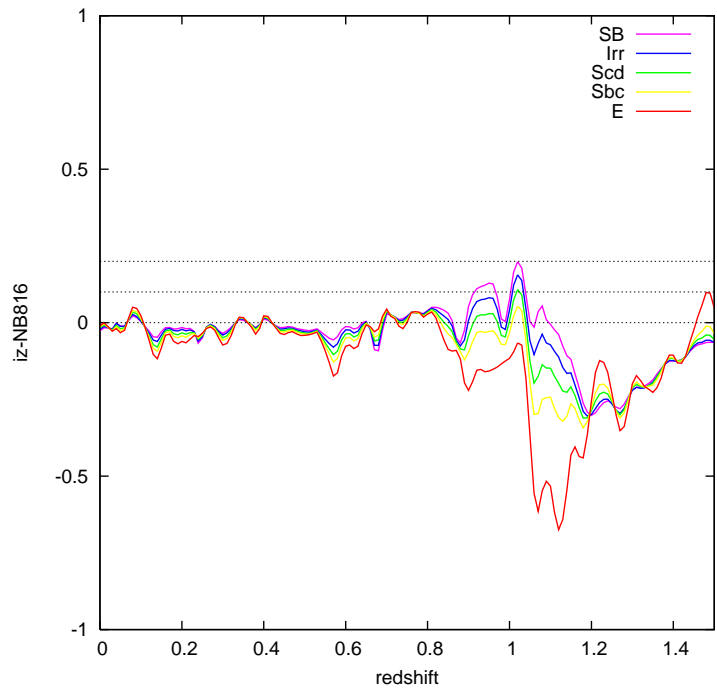}
\caption{Evolution of $iz - \mathit{NB816}$ color as a function of redshift. 
Color loci of model galaxies (SB, Irr, Scd, Sbc, E) are shown with pink, blue, 
green, yellow, and red lines, respectively. 
The horizontal lines corresponds to $iz - \mathit{NB816} = 0.0, 0.1$ and $0.2$. 
Note that emission line features are not included in this diagram. 
\label{iz-nbevo}}
\end{figure}

However, we have to mention why we find galaxies at $z \approx 1.02$ 
in spectroscopy.
To investigate the detection at $z \approx 1.02$, we calculated evolution of 
$iz - {\mathit NB816}$ color 
by using the population synthesis model, GALAXEV (Bruzual \& Charlot 2003), and 
found that continuum feature of galaxies at $z \sim 1.0$ makes {\it NB}816-excess 
as shown in Figure \ref{iz-nbevo}. 
Note that emission-line features are not included in Figure \ref{iz-nbevo}. 
Although one of the \OII\ emitters at $z \sim 1.2$ 
with small {\it NB}816-excess are removed, all four detections at $z \approx 1.02$ can be 
removed by criterion of $iz - {\mathit NB816} \geq 0.2$ as expected. 
Therefore, candidate line-emitting objects are selected with the criterion of 
$iz - {\mathit NB816} \geq 0.2$. 
Thus, it is expected that there is little contamination from objects at other different
redshifts in our \OII\ emitter sample.

In the case of the SDF, 
redshifts of the following four \OII\ emitters selected here
(ID 27396, ID 29922, ID 36025, and ID 178592) were also confirmed 
spectroscopically by Ly et al. (2006). 
Their spectroscopic redshifts are 1.1818, 1.1813, 1.1798, and 1.1783, respectively.

\subsection{\OII\ Luminosity}

The line flux, $f_{\rm L}$, is given by: 
\begin{equation}
f_{\rm L} = \Delta {\rm NB} \frac{f_{\rm NB} - f_{iz}}{1-0.57(\Delta {\rm NB}/\Delta i)},
\end{equation}
where $f_{\rm NB}$ and $f_{iz}$ are the flux densities in each filter, with
$\Delta {\rm NB}$ and $\Delta i$ being the effective bandwidth in those filters. 
The limiting observed line flux in the COSMOS field and that of the SDF are 
$1.4 \times 10^{-17}$ ergs s$^{-1}$ cm$^{-2}$ and $8.7 \times 10^{-18}$ ergs s$^{-1}$ cm$^{-2}$,
respectively.

In order to obtain the [O {\sc ii}] luminosity for each source, 
we apply a mean internal extinction correction to each object. 
To compare our result with the previous investigations compiled by Hopkins (2004), 
we use the same extinction correction method as that proposed by Hopkins (2004); 
$A_{\rm{H}\alpha}=1.0$ mag and the Cardelli et al. (1989) Galactic obscuration curve. 
Note that this corresponds to $A_{\rm{V}}=1.22$ and $A_{\rm[OII]}=1.87$ mag at the 
wavelength of [O {\sc ii}]$\lambda$3727. We also apply a statistical correction 
(21\%; the average value of flux decrease due to the {\it NB}816 filter transmission) 
to the measured flux because the filter transmission function is not square in shape 
(Fujita et al. 2003b). 
The [O {\sc ii}] flux is then given by:
\begin{eqnarray}
f_{\rm{cor}}(\textrm{[O {\sc ii}]} ) = f_{\rm L} \times 
10^{0.4A_{\rm[OII]}} \times 1.21
\end{eqnarray}
Finally, the [O {\sc ii}] luminosity is estimated by $L(\textrm{[O {\sc ii}]}) = 
4\pi d_{\rm{L}}^2 f_{\rm{cor}}(\textrm{[O {\sc ii}]})$. 
In this procedure, we assume that all the \OIIlam\ emitters are located at $z=1.187$ 
that is the redshift corresponding to the central wavelength of {\it NB}816 filter. 
Therefore, the luminosity distance is set to be $d_{\rm L}=8190$ Mpc. 
For the COSMOS field, the aperture effect is corrected by using the offset between $3^{\prime \prime}$ 
diameter aperture and total magnitude given in the COSMOS official photometric catalog
(Capak et al. 2007). For the SDF, we used the total magnitude in 
the SDF official catalog (Kashikawa et al. 2004). 
The basic data of \OII\ emitter candidates, 
including $3^{\prime \prime}$ aperture magnitude of $i^\prime$, ${\it NB}816$, 
$z^\prime$, $iz$, and photometric errors  in the COSMOS field and the SDF are 
listed in Tables \ref{allcosmos} and \ref{allsdf}, respectively. 
These tables also include line flux $f_{\rm L}$ and \OII\ luminosity $L(\textrm{[O {\sc ii}]})$.

\section{DISCUSSION}

\subsection{Luminosity Function of \OII\ Emitters}

\begin{figure}
\begin{center}
\epsscale{1.3}
\plottwo{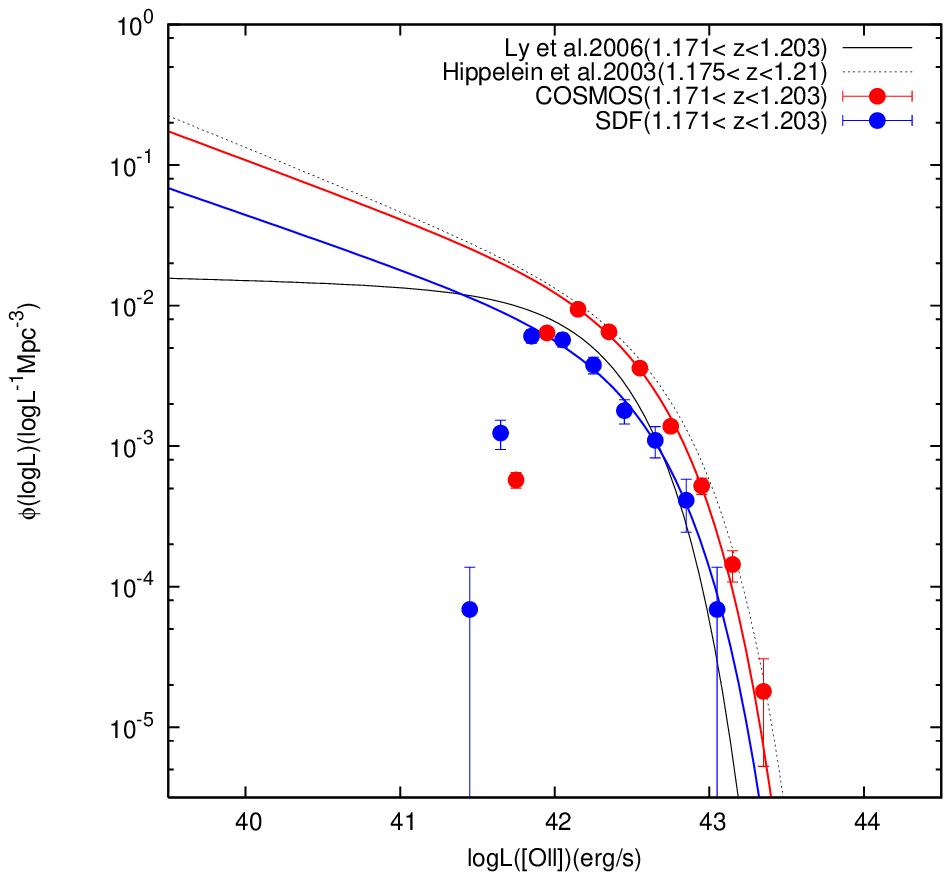}{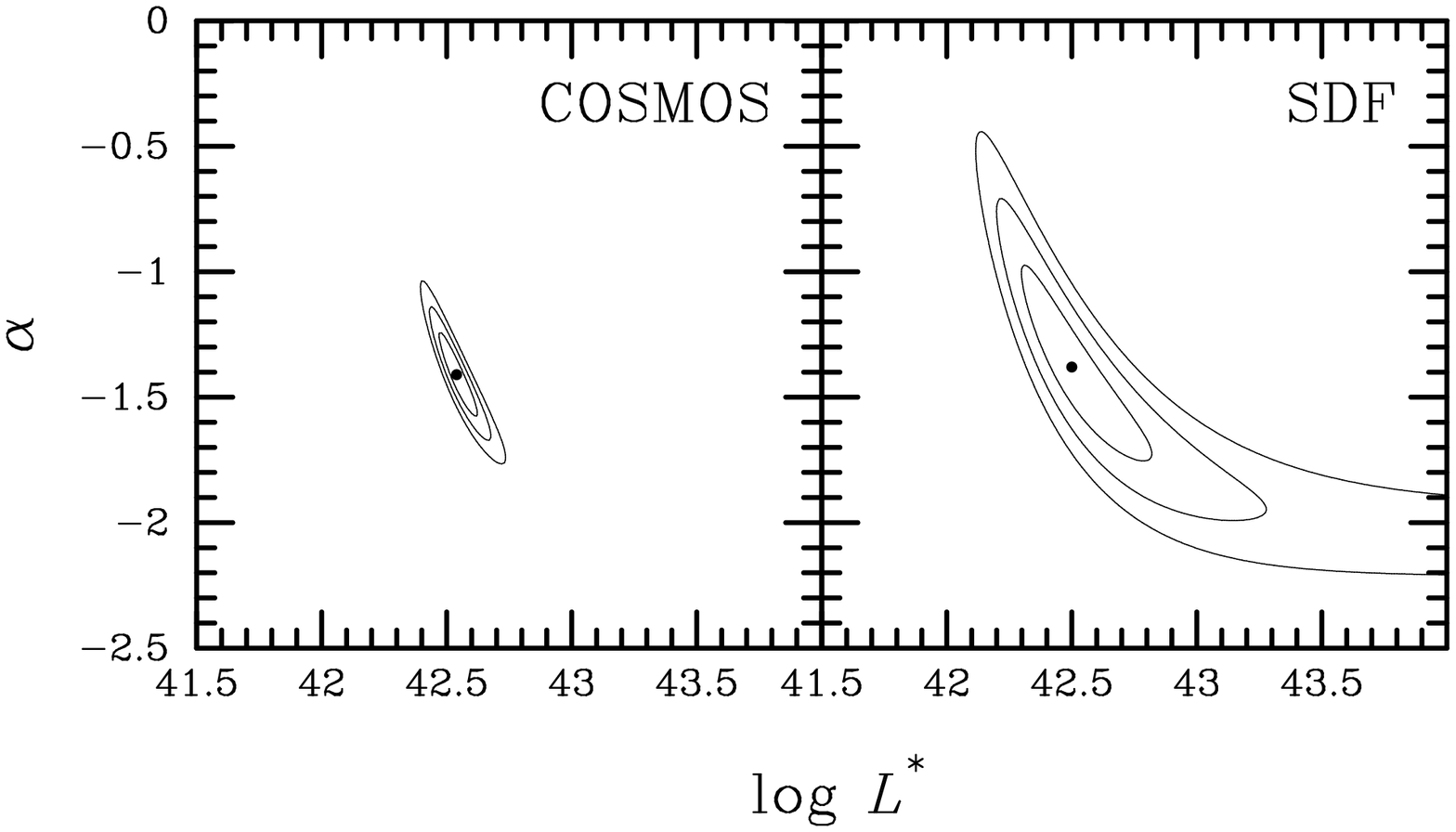}
\caption{$Top$: Luminosity function at $z \approx 1.2$.  
Red points show extinction-corrected LF of the COSMOS field, 
blue points show extinction-corrected LF of the SDF. 
Red and blue lines are the best fitted Schechter function. 
The \OII\ LF derived by Ly et al. (2006) and 
Hippelein et al. (2003) are shown with black line and dotted line, respectively.
$Bottom$: 1, 2, and 3$\sigma$ error contours for the best-fit \OII\ LF parameters of 
the COSMOS field ($left$) and the SDF ($right$). 
\label{LF}}
\end{center}
\end{figure}

In order to investigate the star formation activity in the galaxies at $z \approx 1.2$ detected here, 
we construct the luminosity function (LF) for our COSMOS and SDF \OII\ emitter samples. 
The LF is given by the relation, 
\begin{equation} 
\Phi_i (\log L(\textrm{[O {\sc ii}]} )) \Delta \log L(\textrm{[O {\sc ii}]} ) = \frac{N_i}{V_{\rm c}} , 
\end{equation} 
where $V_{\rm c}$ is the comoving volume and $N_i$ is number of galaxies with \OII\ luminosity within 
the interval of $\log L($\OII $)\pm0.5\Delta\log L($\OII $)$. 
We have used $\Delta\log L($\OII $)$ = 0.2. 
The \OII\ LFs of the COSMOS field and SDF are shown in the upper panel of Figure \ref{LF}. 
The number counts and the LFs obtained in both 
the COSMOS field and SDF are given in Table \ref{LFnumber}. 

The samples that characterize the faint end slope of the \OII\ LF may be 
incomplete due to our selection criteria. 
The \OII\ LF looks to be incomplete at $\log L\sim42.0$ for the COSMOS field 
and at $\log L\sim41.8$ for the SDF.
To investigate the lowest luminosity at which the \OII\ LF is still 
sufficiently complete, we checked the luminosity where the maximum number count appears in the LF.
We examined several cases when the bins of the LF are slightly shifted 
for avoiding that the selection of bins of the LF affects this analysis. 
We finally found $\log L_{\rm lim} = 42.03$ and 41.77, are expected to be 
the completeness limit in the COSMOS field and SDF, respectively. 
We fit the [O {\sc ii}] LFs at $L\ge L_{\rm lim}$ with the Schechter function 
(Schechter 1976) by the maximum likelihood parametric fit (the STY method; Sandage et al. 1979). 
 
We obtain the following best-fitting parameters for our \OII\ emitters 
in the COSMOS field with 
$\alpha = -1.41^{+0.16}_{-0.15}$, 
$\log \phi_* = -2.37^{+0.10}_{-0.12}$, and 
$\log L_* = 42.54^{+0.07}_{-0.06}$, 
in the SDF with 
$\alpha = -1.38^{+0.40}_{-0.37}$, 
$\log \phi_* = -2.67^{+0.28}_{-0.49}$, and 
$\log L_* = 42.50^{+0.32}_{-0.20}$. 
The results of Schechter function fit are overlayed in the upper panel of Figure \ref{LF}. 

Since the Schechter function parameter $\alpha$ is sensitive to the incompleteness 
at the faint end, we also investigate the change of $\alpha$ as a function of 
limiting luminosity $\log L_{\rm{lim}}$. We confirmed that the best-fit $\alpha$
changes fast from when the $\log L_{\rm{lim}}$ is decreased from 42.03 (COSMOS) 
and 41.77 (SDF). 
On the other hand, the best-fit $\alpha$ changes slowly from when the $\log L_{\rm{lim}}$ 
is increased from 42.03 (COSMOS) and 41.77 (SDF).
For example, ($\log L_{\rm{lim}}, \alpha$) = (41.98, -1.15$^{+0.14}_{-0.13}$), 
(42.03, -1.41$^{+0.16}_{-0.15}$), (42.08, -1.55$^{+0.19}_{-0.17}$) for the 
COSMOS field, and ($\log L_{\rm{lim}}, \alpha$) = (41.72, -1.08$^{+0.36}_{-0.37}$), 
(41.77, -1.38$^{+0.40}_{-0.37}$), (41.82, -1.60$^{+0.43}_{-0.41}$) for the SDF, respectively. 
Thus, we conclude that the adopted completeness limits of $\log L_{\rm{lim}}$= 42.03 (COSMOS) 
and 41.77 (SDF) are well determined and that derived Schechter parameters are robust for the incompleteness.

When our results of the COSMOS field and SDF are compared, 
the number density of the COSMOS field is higher by a factor of $\approx 2$ than 
that of the SDF because the $\phi_*$ of the COSMOS field is twice as large as that of the SDF 
while $L_*$ is nearly the same for the both fields. 
In spite of deeper survey depth of the SDF, the faint end slope of the LF 
for the COSMOS field and the SDF yields a similar slope. 
The lower panel of Figure \ref{LF} shows the 1, 2, and 3 $\sigma$ contour levels in the $\alpha - \log L_*$ 
parameter space from the STY analysis. It can be seen that the $\alpha$ and $L_*$ 
are highly correlated. 
We therefore conclude that the large number statics of the COSMOS field provides
both the reliable bright and faint ends of the \OII\ LF.

The \OII\ LF of previous studies (Ly et al. 2006; Hippelein et al. 2003) 
are also plotted in the upper panel of Figure \ref{LF}. 
Hippelein et al. (2003) surveyed an area of 309 arcmin$^2$ with using their own medium-band filters 
and selected 119 \OII\ emitters with the limiting observed luminosity of $\log L($\OII $) = 41.38$ ergs s$^{-1}$. 
Our \OII\ LF of the COSMOS field is similar to that of Hippelein et al. (2003). 
The faint end slope $\alpha = -1.45$ of Hippelein et al. (2003) is in agreement with 
our value, $\alpha = -1.41^{+0.16}_{-0.15}$, in the COSMOS field. 
Ly et al. (2006) selected 894 \OII\ emitters in the SDF by using the same catalogue of our analysis. 
The difference between our LF of the SDF and that of Ly et al. (2006) may be 
attributed to the following different source selection procedures: 
Ly et al. (2006) used
(1) $2^{\prime \prime}$ diameter aperture magnitude,
(2) criterion of $iz - {\mathit{NB}816} > 2.5\sigma_{iz-\mathit{NB}816}$, 
$iz - \mathit{NB816} = 0.25$, and the median for the {\it NB}816-excess of 0.10,
(3) $B - V$ vs. $R_{c} - i^\prime$ and $V - R_{c}$ vs. $i^\prime - z^\prime$ color-color diagrams 
to isolate \OII\ emitters from other emitters at different redshifts, and
(4) SFR-dependent correction of Hopkins et al. (2001) for extinction correction. 

The number density of the \OII\ emitters in the SDF is always lower than those of 
the COSMOS field and Hippelein et al. (2003) at all the concerned luminosities. 
This difference is not due to selection method but probably to the effect of cosmic variance. 
One possible idea seems to be that the SDF samples a relatively low number density region 
of star-forming galaxies at $z \approx 1.2$ because the number density obtained in 
the COSMOS field with the covered area of $\sim 2$ deg$^2$ is expected to be 
closer to an average number density in the Universe. 
Further analysis of \OII\ luminosity function at $z \sim 1.2$ in other fields will be 
required to confirm that this is indeed effect of cosmic variance. 

\subsection{Luminosity Density and Star Formation Rate Density}

The \OII\ luminosity density is obtained by integrating the LF:
\begin{equation}
\mathcal{L}(\textrm{\OII} ) = \int^{\infty}_{0} \Phi(L)LdL = \Gamma(\alpha + 2) \phi_* L_*, 
\end{equation}
where $\Gamma$ is Gamma function. 
We find the total \OII\ luminosity per unit comoving volume to be
$10^{40.35^{+0.08}_{-0.06}}$ 
and $10^{39.99^{+0.22}_{-0.09}}$ ergs s$^{-1}$ Mpc$^{-3}$ at $z \approx 1.2$ 
in the COSMOS field and the SDF from our best fit LF, respectively. 
The star formation rate can be estimated from the \OII\ luminosity using the relation 
\begin{equation}
SFR = 1.41 \times 10^{-41} L(\textrm{[O {\sc ii}]})\:M_\Sun {\rm yr}^{-1},
\end{equation} 
where $L($\OII $)$ is in units of ergs s$^{-1}$ (Kennicutt 1998). 
Thus, the \OII\ luminosity density can be translated into the SFR density of 
$\rho_{\rm SFR} \simeq 0.32^{+0.06}_{-0.04}$ and $0.14^{+0.09}_{-0.03}$ $M_\Sun$ yr$^{-1}$ Mpc$^{-3}$ 
in the COSMOS field and the SDF, respectively. 
This SFRD from \OII\ luminosity in the COSMOS field is estimated by using the statistically 
largest sample studied so far. The vertical error bar due to the 
number statistics is definitively small, and the horizontal error bar due to the width 
of the narrowband filter is also small compared with other spectroscopic surveys.

\begin{figure}
\begin{center}
\epsscale{1.0}
\plotone{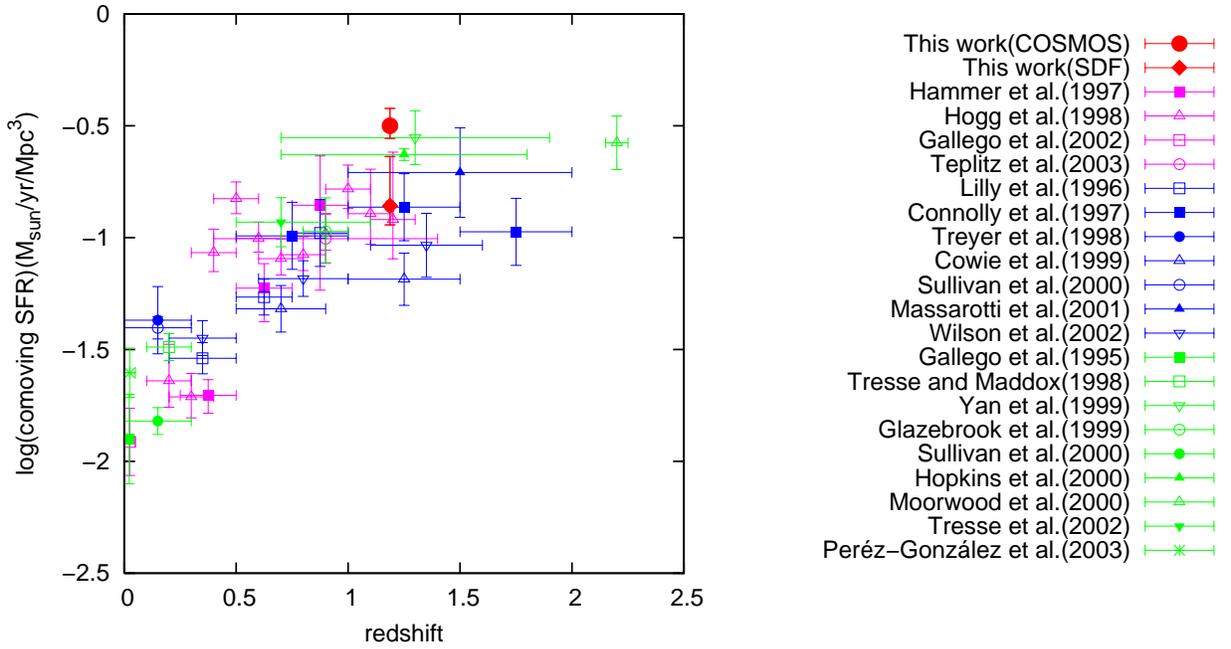}
\caption{Star formation rate density (SFRD) at $z \approx 1.2$ derived from 
this study (large red filled circle for the COSMOS field, diamond for the SDF) 
shown together with the previous investigations compiled by Hopkins (2004). 
SFRD estimated from H$\alpha$, \OII\, and UV continuum are shown 
as green points (P\'erez-Gonz\'alez et al. 2003; Tresse et al. 2002; 
Moorwood et al. 2000; Hopkins et al. 2000; Sullivan et al. 2000; Glazebrook et al. 1999; 
Yan et al. 1999; Tresse \& Madox 1998; Gallego et al. 1995), 
pink points (Teplitz et al. 2003; Gallego et al. 2002; Hogg et al. 1998; 
Hammer et al. 1997), 
and blue points (Wilson et al. 2002; Massarotti et al. 2001; 
Sullivan et al. 2000; Cowie et al. 1999; Treyer et al. 1998; Connolly et al. 1997; Lilly et al. 1996). 
\label{Ha:MadauPlot}}
\end{center}
\end{figure}

Figure \ref{Ha:MadauPlot} shows the evolution of the SFRD as a function of 
redshift from $z = 2.5$ to $z = 0$. 
The previous investigations plotted here are compiled by Hopkins (2004), 
which have been converted to a common cosmology, consistent SFR calibrations, and 
consistent dust obscuration corrections where necessary. 
Our results follow the general of the strong decrease in the SFRD from $z \sim 1$ to $z = 0$. 
If we carefully compare our results with previous ones, we find that
the SFRD measured in the COSMOS field is consistent with that of Yan et al. (1999), while 
that of the SDF is consistent with that of Hogg et al. (1998). 
The difference among the SFRDs of the COSMOS field, SDF,
and the other surveys may not surprising because the SFRD 
varies from space to space within at least factor of two due to the cosmic variance. 

\subsection{Spatial Distribution and Angular Two-Point Correlation Function}

\begin{figure}
\begin{center}
\epsscale{0.6}
\plotone{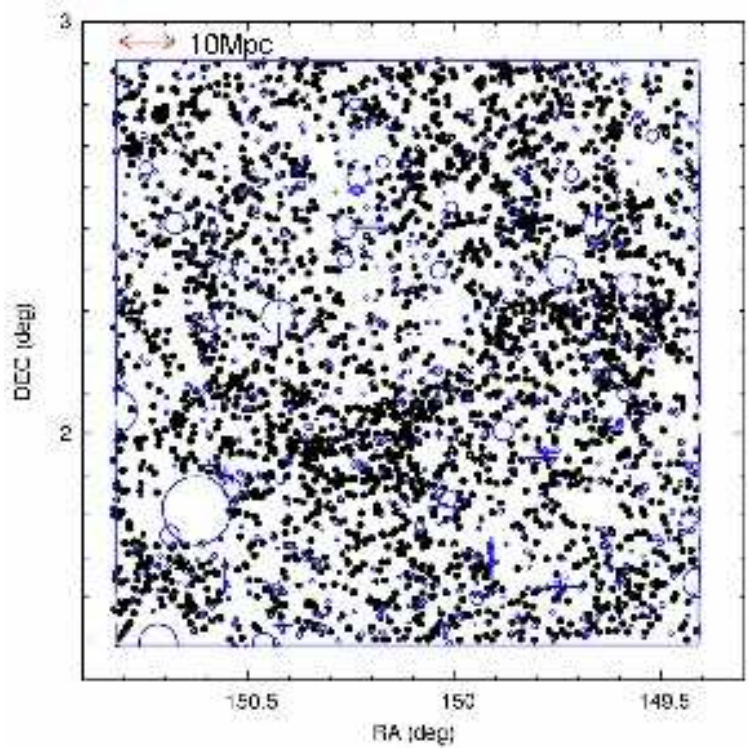}
\caption{Spatial distributions of our \OII\ emitter candidates 
in the COSMOS field (black points). 
The areas with blue circles are  masked out for the detection. 
The red arrow in the upper-left corner shows the co-moving scale of 10 Mpc.
\label{Ha:RaDec}}
\end{center}
\end{figure}

Figure \ref{Ha:RaDec} shows the spatial distribution of our 3176 \OII\ emitter 
candidates in the COSMOS field. 
There can be seen some clustering regions. In particular, it is worthwhile
noting that there appears to be a filamentary structure near the central
region of the field in  Figure \ref{Ha:RaDec}. If this is real, this filament
extends across $\sim 40$ Mpc. 
At $z \sim 1$, large-scale clustering of red galaxies with 
the scale of $> 10$ Mpc have been identified 
(e.g., Tanaka et al. 2001; Nakata et al. 2005; Gal, Lubin \& Squires 2005). 
Our result suggests that not only red galaxies but also relatively blue, 
star-forming galaxies form large-scale clustering at these redshifts. 

\begin{figure}
\begin{center}
\epsscale{1.3}
\plottwo{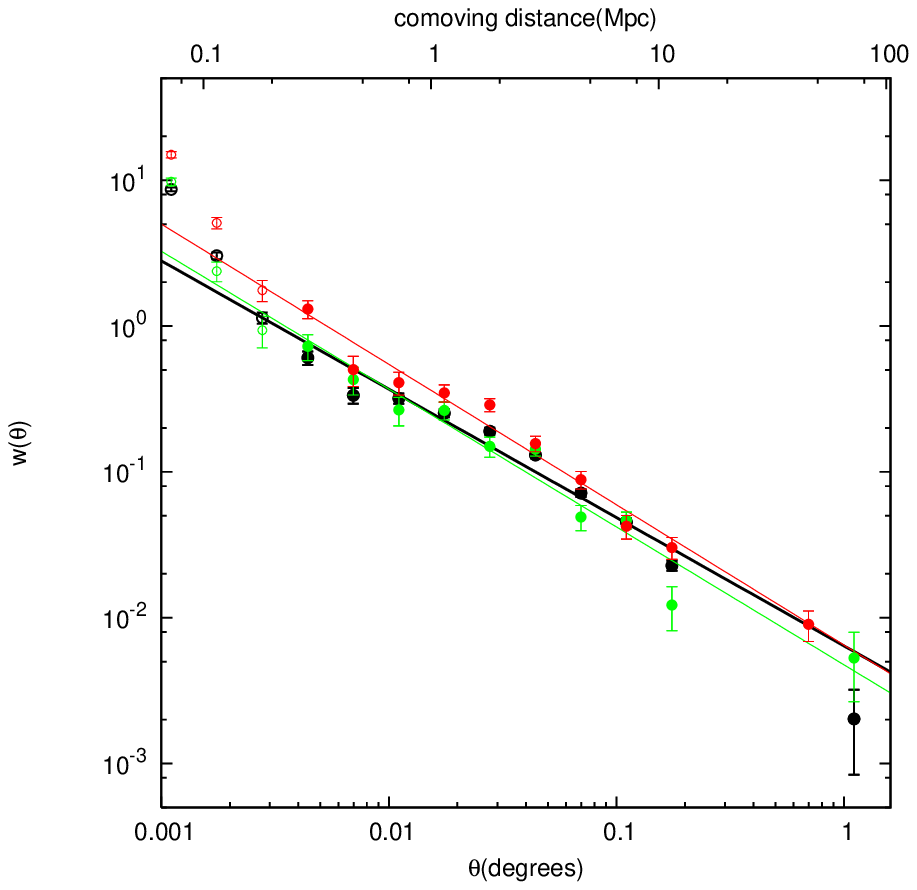}{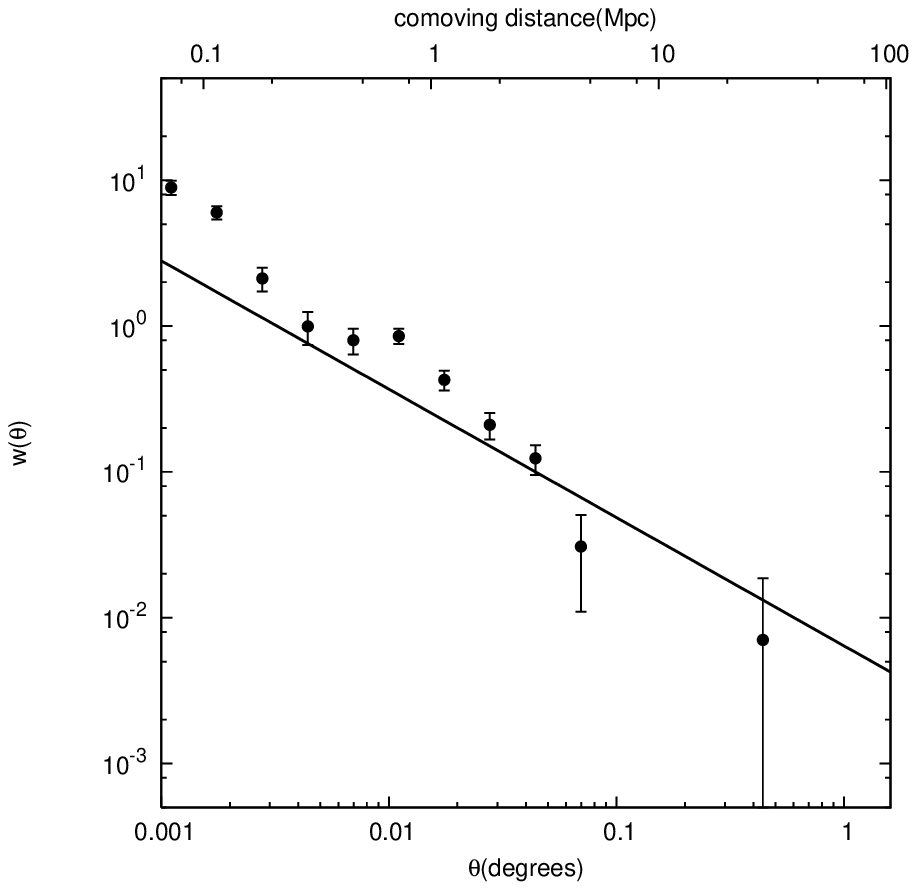}
\caption{Angular two-point correlation function of the \OII\ emitter candidates 
at $z \approx 1.2$ in the COSMOS field ($Top$) and in the SDF ($Bottom$). 
Red and green points in $Top$ shows the ACF of bright \OII\ emitter candidates 
($42.3 \le \log L($\OII $)$) and that of faint \OII\ emitter candidates 
($42.03 \le \log L($\OII $) < 42.3$) in the COSMOS field, respectively.  
Black, red, and green lines show the best-fit power law for the whole, bright, and faint samples 
of the COSMOS \OII\ emitters for a range of $0.004 < \theta < 1.11$ (filled circles), respectively. 
The scale on the top axis denotes the projected co-moving distance at $z \approx 1.2$. 
\label{Ha:ACF}}
\end{center}
\end{figure}

To discuss the clustering properties more quantitatively, 
we derive the angular two-point correlation function (ACF), $w(\theta)$,
using the estimator defined by Landy \& Szalay (1993),
\begin{equation}
 w(\theta) = \frac{DD(\theta)-2DR(\theta)+RR(\theta)}{RR(\theta)},
 \label{two-point}
\end{equation}
where $DD(\theta)$, $DR(\theta)$, and $RR(\theta)$ are normalized numbers of
galaxy-galaxy, galaxy-random and random-random pairs, respectively. 
The random sample consists of 100,000 sources with the same geometrical
constraints as the galaxy sample. 
Figure \ref{Ha:ACF} shows the ACF of our 3176 and 294 \OII\ emitter 
candidates in the COSMOS field ($Top$) and SDF ($Bottom$), respectively.
The wide area of the COSMOS allows us to see the clustering properties at 
$z \sim 1.2$, $\sim$ 90 $\times$ 90 Mpc$^2$ (comoving), 
while that of the SDF is $\sim$ 30 $\times$ 40 Mpc$^2$. 
Note that our narrowband survey samples the thickness of $\sim$ 70 Mpc at $z \sim 1.2$ 
along the line of sight, corresponding to the band width of the {\it NB}816.

In spite of the different field, amplitude and slope of the ACF are almost 
consistent among the COSMOS field and the SDF. 
The ACFs of \OII\ emitters show significant excess on small scales. 
For sample of $0.004 < \theta < 1.11$ degree (filled black points), 
the ACF of the COSMOS field is fit by power law, 
$w(\theta) = A_w \theta^\beta$; 
$A_w = 0.0064 \pm 0.0014$ and $\beta = -0.88 \pm 0.06$ (black line). 
Note we did not include negative value of $w(\theta)$ on 
$0.2 < \theta < 1$. 

If the real-space correlation function $\xi (r)$ is modeled as a power law, 
$\xi (r) = (r / r_0)^{- \gamma}$, 
we can infer the best-fit power law for $\xi (r)$ from $w(\theta)$ using the Limber transform 
(Peebles 1980; Phillipps 2005). 
This gives
\begin{equation}
\gamma = 1 - \beta,
\end{equation}  
\begin{equation}
 w(\theta) = I_\gamma S(\Psi ) r^{\gamma }_{0} \theta^{1 - \gamma },
\end{equation}
where 
\begin{equation}
 I_\gamma = \frac{\sqrt{\pi}\Gamma(\gamma / 2 - 1/ 2)}{\Gamma (\gamma / 2)},
\end{equation}
and 
\begin{equation}
S(\Psi ) = \frac{\int \Psi  ^2 x^{5 - \gamma }dx}{(\int \Psi x^2 dx)^2},
\end{equation}
where $x$ is the proper distance. Note that $S(\Psi )$ depends only on the selection function $\Psi$. 
Assuming that the redshift distribution of the \OII\ emitters is a top-hat shape of $z = 1.187\pm 0.016$, 
we convert the amplitude $A_w$ to the correlation length $r_0$.
Then we obtain $\gamma = 1.88 \pm 0.06$ and $r_0 = 1.67^{+ 0.45}_{- 0.41}$ $h^{-1}$ Mpc, 
where $h = H_0/100 \; {\rm km \; s^{-1} \; Mpc^{-1}}$.

The correlation function at $z \sim 1$ is also studied by Le F$\grave{{\rm e}}$vre et al. 
(2005) and Coil et al. (2006). Le F$\grave{{\rm e}}$vre et al. (2005) analysed 7155 galaxies 
with $17.5 < I_{AB} < 24$ at $0.2 < z \leq 2.1$, from the VIMOS VLT Deep Survey (VVDS). 
They obtained $\gamma = 1.96^{+ 0.27}_{- 0.21}$ and 
$r_0 = 3.09^{+ 0.61}_{- 0.65}$ $h^{-1}$ Mpc, for the sub-sample of 561 galaxies at $1.1 < z < 1.3$ 
by  power law fitting of the ACF in a proper length range of 0.1 -- 10 $h^{-1}$ Mpc. 
Coil et al. (2006) analysed $\sim$25,000 galaxies with $M_{\rm B} < -19$ at $0.7 < z < 1.3$ 
from the DEEP2 Galaxy Redshift Survey. From power law fitting of the ACF at 0.1 -- 20 $h^{-1}$ Mpc, 
$\gamma = 1.71 \pm 0.03$ and $r_0 = 3.69 \pm 0.14$ $h^{-1}$ Mpc are obtained for the DEEP2 sub-sample 
(10530 galaxies with $M_{\rm B} < -19$).

The slope of the COSMOS is consistent with that of the VVDS within the errors. 
The relatively small value of $\gamma = 1.71$ is obtained in the DEEP2 ($0.7 < z < 1.3$) 
with respect to those of the COSMOS ($z \approx 1.2$) and VVDS ($1.1 < z < 1.3$) samples. 
This is consistent with the result from the VVDS, which shows slight increase of the 
slope $\gamma$ with increasing redshift (Le F$\grave{{\rm e}}$vre et al. 2005). 

The correlation length of the COSMOS is smaller than those of the VVDS and DEEP2. 
This difference may be due to that the brighter galaxies are selected by the VVDS 
($-23 < M_{\rm B} < -19$) and DEEP2 ($M_{\rm B} < -19$) than by our COSMOS 
\OII\ sample ($M_{\rm B} < -18$ estimated by $z^\prime$ magnitude). 
It has been extensively studied that more luminous galaxies are more strongly 
clustered in the local universe (Norberg et al. 2001, 2002; Zehavi et al. 2005) 
and at $z \sim 1$ (Coil et al. 2006; Pollo et al. 2006). 
Further, the selection method adopted in the DEEP2 tends to exclude relatively blue galaxies 
and may result in the larger correlation length than those of the COSMOS \OII\ sample and the VVDS sample. 
It is known that the early type galaxies tend to show stronger clustering than 
late type ones (Loveday, Tresse \& Maddox 1999; Norberg et al. 2002).

Since our samples are selected by emission line features and thus 
biased to star-forming galaxies, the comparison is not straightforward. 
Therefore, we also compare our results with those of Meneux et al. (2006). 
They measure the spectral type dependence of correlation function using a 
sample of 6495 VVDS galaxies. The galaxies with spectroscopic redshifts 
were divided into four spectral classes from E/S0 (type 1), early spiral, 
late spiral and irregular/star-forming galaxies (type 4). 
Our \OII\ samples are expected to be almost late spiral and irregular star-forming galaxies 
(see, Figures \ref{Ha:BVrcolor} and \ref{Ha:Brizcolor}), we compare with merging type 3 and 4. 
They find that merging type 3 and 4 galaxies at $0.9 < z < 1.2$ (1030 galaxies, effective redshift = 1.032) 
have $\gamma = 1.86^{+ 0.11}_{- 0.08}$ and $r_0 = 2.58^{+ 0.25}_{- 0.22}$ $h^{-1}$ Mpc. 
The slope is in good agreement with that of the COSMOS within the errors. 
Although this correlation length is smaller than that of all galaxies in the VVDS and DEEP2, 
and closer to our value $r_0 = 1.67$ $h^{-1}$ Mpc, it is still larger than our result. 
Note that the magnitude range of both analyses is nearly the same. 
A interpretation of this result is that strong emission line galaxies show weaker 
clustering than overall late type galaxies.

To investigate the dependence of \OII\ luminosity, we also derive the ACFs for sub samples of 
different luminosity bins in the COSMOS field;
(1) $42.03 \le \log L($\OII $) < 42.30$, and (2) $42.30 \le \log L($\OII $)$. 
Note that $\log L($\OII $)= 42.03$ corresponds to the luminosity limit at the faint end. 
The green and red points in the upper panel of Figure \ref{Ha:ACF} show the ACFs 
obtained for the faint  subsample and for the bright subsample, respectively. 
By power-law fitting of each ACF in a range of $0.004 < \theta < 1.11$ degree 
we obtain $A_w = 0.0048 \pm 0.0015$ and  $\beta = -0.95 \pm 0.09$ (shown by the green line in Figure 9) 
or $\gamma = 1.95 \pm 0.09$ and $r_0 = 1.65^{+ 0.60}_{- 0.55}$ $h^{-1}$ Mpc for the faint sample, 
and $A_w = 0.0065 \pm 0.0014$ and $\beta = -0.96 \pm 0.06$ (shown by the red line in Figure 9) 
or $\gamma = 1.96 \pm 0.06$ and $r_0 = 2.00^{+ 0.45}_{- 0.43}$ $h^{-1}$ Mpc for the bright sample. 
The ACF amplitudes in the COSMOS field are given in Table \ref{acfdata}. 

Our result that clustering increases with increasing luminosity is consistent with many previous investigations 
(Norberg et al. 2002; Zehavi et al. 2005; Coil et al. 2006; Lee et al. 2006). 
The deviation from the power law at a small scale is also reported by previous investigations 
(Zehavi et al. 2004; Ouchi et al. 2005b; Coil et al. 2006; Kashikawa et al. 2006; Lee et al. 2006). 
This could be explained by galaxy multiplicity in a single dark matter halo at higher luminosities and on smaller scale. 
We do not analyse luminosity dependence of the ACF in the SDF, because compared with the COSMOS field, 
the number statistics of the SDF is too small to carry out this detailed analysis of the clustering property.  

\section{SUMMARY}

A wide-field narrowband imaging survey for \OIIlam\ emitting galaxies 
at $z \approx 1.2$ in the HST COSMOS 2 square degree field has been presented. 
Our main results and conclusions are as follows.

1. In this survey, we have found 3176 \OII\ emitting galaxies
in a comoving volume of 5.56$\times 10^5$ Mpc$^3$. 
This is the largest survey for \OII\ emitters beyond $z = 1 $ to date.

2. Although our survey samples numerous \OII\ emitting galaxies,
most of them are relatively brighter ones with log $L$([O {\sc ii}])
$> 10^{42.03}$ erg s$^{-1}$. Therefore, it is difficult to investigate
fainter parts of [O {\sc ii}] luminosity function. In order to 
obtain more reliable information, we have also analysed optical
photometric data of the Subaru Deep Field (SDF) available for
public (Kashikawa et al. 2004). Then we have constructed the \OII\ luminosity function
and found the star formation rate density of $0.32^{+0.06}_{-0.04}$ $M_{\sun}$ yr$^{-1}$ Mpc$^{-3}$ 
in the COSMOS field. This result is consistent with those of previous works
at similar redshifts (see Hopkins 2004) within the errors. 

3. The spatial distributions of \OII\ emitters show some over density regions
over the entire field. In particular, a filamentary structure is found in the
central region of the COSMOS field. Its extension is estimated as several tens Mpc. 
These results suggest that the star-forming galaxies at $z \approx 1.2$ found here
tend to cluster.

4. Our angular two-point correlation function of the COSMOS field is 
fit by a power law, $w(\theta) = (0.0064 \pm 0.0014) \theta^{-0.88 \pm 0.06}$. 
We confirmed the previous results that more luminous galaxies have stronger clustering 
strength. We found significant deviation from power law at small scale in the ACF of \OII\ emitters.
This result is consistent with previous surveys (Zehavi et al. 2004; 
Ouchi et al. 2005; Kashikawa et al. 2006), 
which suggests that \OII\ emitters at $z \sim 1.2$ are biased tracers of the mass density field.

The HST COSMOS treasury program was supported through NASA grant HST-GO-09822. 
We greatly acknowledge the contributions of the entire COSMOS collaboration 
consisting of more than 70 scientists. The COSMOS science meeting in May 2005 
was supported by in part by the NSF through grant OISE-0456439. 
We would also like to thank the Subaru Telescope staff for their invaluable help. 
This work was financially supported in part by the JSPS (Nos. 15340059 and 17253001). 
SSS and TN are JSPS fellows.

\appendix

\section*{APPENDIX}

\begin{figure}
\begin{center}
\epsscale{1.0}
\plotone{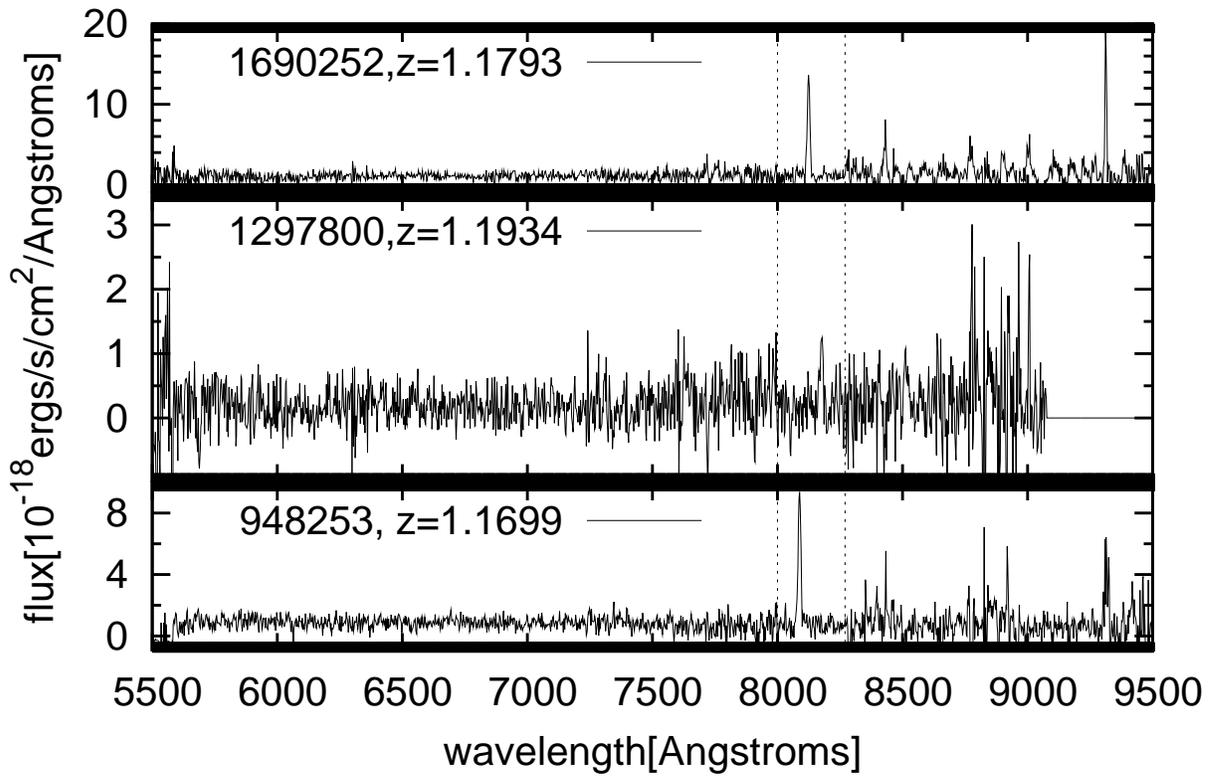}
\caption{Optical spectra of COSMOS \OII\ emitter candidates. 
Their ID numbers and spectroscopic redshifts are shown at left top in each panel. 
The vertical dotted line shows the wavelength coverage of {\it NB}816. 
They are all in the expected redshift range, $1.17 < z < 1.2$. 
\label{spec}}
\end{center}
\end{figure}

In this Appendix, we show optical spectra of the following three \OII\ emitters; 
ID 948253, ID 1297800, and ID 1690252. The spectra shown in Figure \ref{spec} were
obtained during the zCOSMOS observing runs with VIMOS on the Very Large 
Telescope (see for details, Lilly et al. 2006). 

\clearpage

\begin{deluxetable}{lcccccccccc}
\tabletypesize{\scriptsize}
\tablenum{1}
\tablecaption{\label{oii:tab:sdfcosmos}Basic properties of the COSMOS and SDF}
\tablewidth{0pt}
\tablehead{
\colhead{Field} &
\colhead{Covered sky area} &
\colhead{Survey volume} &
\colhead{} &
\colhead{} &
\colhead{limiting} &
\colhead{magnitude} &
\colhead{} &
\colhead{} &
\colhead{} \\ 
\colhead{} &
\colhead{(arcmin$^2$)} &
\colhead{($10^{4}$Mpc$^3$)}  &
\colhead{$B$\tablenotemark} &
\colhead{$V$\tablenotemark} &
\colhead{$r^\prime$\tablenotemark} &
\colhead{$R_c$\tablenotemark} &
\colhead{$i^\prime$\tablenotemark} &
\colhead{$z^\prime$\tablenotemark} &
\colhead{{\it NB}816\tablenotemark} &
}
\startdata
COSMOS & 6700  & 55.6 & 27.4  & 26.5 & 26.6 &  -  & 26.1  & 25.4  & 25.6  \\
SDF    & 875   & 7.27  & 27.6  & 27.0  &  -  & 27.2  & 27.0  & 26.2  & 26.1  \\
\enddata
\end{deluxetable}
\begin{deluxetable}{lccc}
\tablenum{2}
\tablecaption{\label{Ha:tab:cover}Emission lines potentially detected inside the
narrowband}
\tablewidth{0pt}
\tablehead{
\colhead{Line} &
\colhead{Redshift range} &
\colhead{$\bar{z}$\tablenotemark{a}} &
\colhead{$d_{\rm L}$\tablenotemark{b}} \\
\colhead{} &
\colhead{$z_1\leq z \leq z_2$} &
\colhead{} &
\colhead{(Mpc)}
}
\startdata
\HA\  & 0.233~~~0.251 & 0.242 & 1220 \\
\OIII & 0.616~~~0.656 & 0.636 & 3800 \\
\HB\  & 0.664~~~0.689 & 0.677 & 4100 \\
\OII  & 1.171~~~1.203 & 1.187 & 8190 \\
\enddata
\tablenotetext{a}{Mean redshift.}
\tablenotetext{b}{{Luminosity distance.}}
\end{deluxetable}
\begin{deluxetable}{ccccccccccccc}
\tabletypesize{\scriptsize}
\rotate  
\tablenum{3}
\tablecolumns{13} 
\tablewidth{0pt} 
\tablecaption{\label{allcosmos}
Photometric properties of the \OII\ emitter candidates in the COSMOS field} 
\tablehead{ 
\colhead{No.} & \colhead{ID}  &  \colhead{RA(J2000)} & \colhead{DEC(J2000)}  &  
\colhead{$i$\tablenotemark{a}}  & \colhead{$\Delta i$\tablenotemark{b}} & 
\colhead{${\it NB}816$\tablenotemark{a}} & \colhead{$\Delta {\it NB}816$\tablenotemark{b}} & 
\colhead{$z^\prime$\tablenotemark{a}}  &  \colhead{$\Delta z^\prime$\tablenotemark{b}} & 
\colhead{$iz$} & \colhead{$\log f_{\rm L}$\tablenotemark{c}} & \colhead{$\log L([{\rm OII}])$\tablenotemark{d}} \\ 
\colhead{} & \colhead{}  &  \colhead{(deg)} & \colhead{(deg)}  & 
\colhead{(mag)}  & \colhead{(mag)}   & 
\colhead{(mag)}  & \colhead{(mag)}   & 
\colhead{(mag)}  & \colhead{(mag)}   &  
\colhead{(mag)}  & \colhead{(ergs s$^{-1}$ cm$^{-2}$)} & \colhead{(ergs s$^{-1})$}  
}
\startdata
1  &   27025  &   150.77573  &    1.622925  &     23.80  &      0.04  &     23.11  &      0.03  &     23.43  &     0.05  &    23.63  &    -16.33  &     42.41  \\
2  &   27180  &   150.74086  &    1.622155  &     24.44  &      0.06  &     24.00  &      0.07  &     24.32  &     0.10  &    24.39  &    -16.80  &     41.94  \\
3  &   28182  &   150.75733  &    1.615727  &     24.10  &      0.05  &     23.63  &      0.05  &     23.66  &     0.06  &    23.89  &    -16.75  &     41.99  \\
4  &   32985  &   150.78337  &    1.592096  &     24.58  &      0.07  &     23.94  &      0.07  &     24.20  &     0.10  &    24.40  &    -16.70  &     42.04  \\
5  &   33031  &   150.76678  &    1.591625  &     24.10  &      0.05  &     23.46  &      0.04  &     23.80  &     0.07  &    23.96  &    -16.49  &     42.25  \\
\enddata
\tablenotetext{a}{3$^{\prime \prime}$ aperture magnitude.}
\tablenotetext{b}{Photometric error.}
\tablenotetext{c}{{Line flux.}}
\tablenotetext{d}{Extinction corrected \OII\ luminosity.}
The complete version of this table is in the electronic edition of
the Journal. The printed edition contains only a sample.
\end{deluxetable}
\begin{deluxetable}{ccccccccccccc}
\tabletypesize{\scriptsize}
\rotate  
\tablenum{4}
\tablecolumns{13} 
\tablewidth{0pt} 
\tablecaption{\label{allsdf}
Photometric properties of the \OII\ emitter candidates in the SDF} 
\tablehead{ 
\colhead{No.} & \colhead{ID}  &  \colhead{RA(J2000)} & \colhead{DEC(J2000)}  &  
\colhead{$i$\tablenotemark{a}}  & \colhead{$\Delta i$\tablenotemark{b}} & 
\colhead{${\it NB}816$\tablenotemark{a}} & \colhead{$\Delta {\it NB}816$\tablenotemark{b}} & 
\colhead{$z^\prime$\tablenotemark{a}}  &  \colhead{$\Delta z^\prime$\tablenotemark{b}} & 
\colhead{$iz$} & \colhead{$\log f_{\rm L}$\tablenotemark{c}} & \colhead{$\log L([{\rm OII}])$\tablenotemark{d}} \\ 
\colhead{} & \colhead{}  &  \colhead{(deg)} & \colhead{(deg)}  & 
\colhead{(mag)}  & \colhead{(mag)}   & 
\colhead{(mag)}  & \colhead{(mag)}   & 
\colhead{(mag)}  & \colhead{(mag)}   &  
\colhead{(mag)}  & \colhead{(ergs s$^{-1}$ cm$^{-2}$)} & \colhead{(ergs s$^{-1})$}  
}
\startdata
    1  &   10459  &   201.35980  &   27.211346  &     24.20  &      0.01  &     23.69  &      0.01  &     23.77  &      0.01  &     24.00  &    -16.73  &     42.01  \\
    2  &   10789  &   201.24632  &   27.210946  &     24.28  &      0.01  &     23.64  &      0.01  &     23.61  &      0.01  &     23.94  &    -16.69  &     42.05  \\
    3  &   10960  &   201.19371  &   27.210909  &     24.13  &      0.01  &     23.51  &      0.01  &     23.57  &      0.01  &     23.85  &    -16.60  &     42.13  \\
    4  &   11651  &   201.15106  &   27.212800  &     24.85  &      0.01  &     24.25  &      0.02  &     24.38  &      0.02  &     24.62  &    -16.88  &     41.85  \\
    5  &   11959  &   201.25951  &   27.213563  &     24.30  &      0.01  &     23.73  &      0.01  &     23.58  &      0.01  &     23.93  &    -16.84  &     41.89  \\
\enddata
\tablenotetext{a}{3$^{\prime \prime}$ aperture magnitude.}
\tablenotetext{b}{Photometric error.}
\tablenotetext{c}{{Line flux.}}
\tablenotetext{d}{Extinction corrected \OII\ luminosity.}
The complete version of this table is in the electronic edition of
the Journal. The printed edition contains only a sample.
\end{deluxetable}
\begin{deluxetable}{cccccccccccc}
\tabletypesize{\scriptsize} 
\rotate
\tablenum{5}
\tablecolumns{12} 
\tablewidth{0pc} 
\tablecaption{\label{LFnumber}\OIIlam\ luminosity function for the COSMOS field and the SDF} 
\tablehead{ 
\colhead{}  &  \multicolumn{5}{c}{COSMOS} &  \colhead{} & \multicolumn{5}{c}{SDF} \\
\cline{2-6} \cline{8-12} \\
\colhead{}  &  \multicolumn{2}{c}{Observed luminosity function} &  \colhead{}  & 
\multicolumn{2}{c}{Corrected luminosity function} 
& \colhead{} & 
\multicolumn{2}{c}{Observed luminosity function}  &  \colhead{}  & 
\multicolumn{2}{c}{Corrected luminosity function} \\ 
\cline{2-3} \cline{5-6} \cline{8-9} \cline{11-12} \\ 
\colhead{$\log L([{\rm OII}])$} & \colhead{$\log \phi$}  & \colhead{Number of}  & 
\colhead{}   & \colhead{$\log \phi$}   & \colhead{Number of}  & \colhead{} & 
\colhead{$\log \phi$}   & \colhead{Number of}    & 
\colhead{}   & \colhead{$\log \phi$}   & \colhead{Number of}  \\
\colhead{(ergs s$^{-1})$} & \colhead{($\log L^{-1}$ Mpc$^{-3}$)}   & \colhead{galaxies}   & 
\colhead{}  & \colhead{($\log L^{-1}$ Mpc$^{-3}$)}    & \colhead{galaxies} & \colhead{}   &
\colhead{($\log L^{-1}$ Mpc$^{-3}$)}   & \colhead{galaxies}   & 
\colhead{}  & \colhead{($\log L^{-1}$ Mpc$^{-3}$)}  & \colhead{galaxies}}
\startdata
40.75 	&	\nodata	 &	\nodata	 &  &	\nodata	 &	\nodata	 &  &	-3.86 	&	2	&  &	\nodata	 &	\nodata	\\
40.95 	&	-3.61	&	27	&  &	\nodata	 &	\nodata	 &  &	-2.62 	&	35	&  &	\nodata	 &	\nodata	\\
41.15 	&	-2.37	&	470	&  &	\nodata	 &	\nodata	 &  &	-2.22 	&	87	&  &	\nodata	 &	\nodata	\\
41.35 	&	-2.01 	&	1080  &  &	\nodata	 &	\nodata	 &  &	-2.23   &	 86 	&  & \nodata &	\nodata	\\
41.45 	&	\nodata	 &	\nodata	 &  &	\nodata	 &	\nodata	 &  &	\nodata	 &	\nodata	 &  &	-4.16 	&	1	\\
41.55 	&	-2.14 	&	810	&  &	\nodata	 &	\nodata	 &   &	-2.52 	&	44	&  &	\nodata	 &	\nodata	\\
41.65 	&	\nodata	 &	\nodata	 &  &	\nodata	 &	\nodata	 &  &	\nodata	 &	\nodata	 &  &	-2.91 	&	18	\\
41.75 	&	-2.37 	&	474	&  &	-3.24 	&	64	&  &	-2.80 	&	23	&  &	\nodata	&	\nodata	\\
41.85 	&	\nodata	 &	\nodata	 &  &	\nodata	 &	\nodata	 &  &	\nodata	 &	\nodata	 &  &	-2.22 	&	88	\\
41.95 	&	-2.72 	&	212	&  &	-2.19 	&	712	&  &	-3.08 	&	12 	&  &	\nodata	&	\nodata	\\
42.05 	&	\nodata	 &	\nodata	 &  &	\nodata	 &	\nodata	 &  &	\nodata	 &	\nodata	 &  &	-2.24 	&	83	\\
42.15 	&	-3.20 	&	70	&  &	-2.03 	&	1048	&  &	-3.56 	&	4 	&  &	\nodata	&	\nodata	\\
42.25 	&	\nodata	 &	\nodata	 &  &	\nodata	 &	\nodata	 &   &	\nodata	 &	\nodata	 &  &	-2.42 	&	55	\\
42.35 	&	-3.55 	&	31	&  &	-2.19 	&	723	&  &	-4.16	&	1	&  &	\nodata	&	\nodata	\\
42.45 	&	\nodata	 &	\nodata	 &  &	\nodata	 &	\nodata	 &  &	\nodata	 &	\nodata	 &  &	-2.75 	&	26	\\
42.55 	&	-4.75 	&	2	&  &	-2.45 	&	399	&  &	\nodata	&	\nodata	&  &	\nodata	&	\nodata	\\
42.65 	&	\nodata	 &	\nodata	 &  &	\nodata	 &	\nodata	 &  &	\nodata	 &	\nodata	 &  &	-2.96 	&	16	\\
42.75 	&	\nodata	 &	\nodata	 &  &	-2.86 	&	154	&  &	\nodata	 &	\nodata	 &  &	\nodata	 &	\nodata	\\
42.85 	&	\nodata	 &	\nodata	 &  &	\nodata	 &	\nodata	 &  &	\nodata	 &	\nodata	 &  &	-3.38 	&	6	\\
42.95 	&	\nodata	 &	\nodata	 &  &	-3.28 	&	58	&  &	\nodata	 &	\nodata	 &  &	\nodata	 &	\nodata	\\
43.05 	&	\nodata	 &	\nodata	 &  &	\nodata	 &	\nodata	 &  &	\nodata	 &	\nodata	 &  &	-4.16 	&	1	\\
43.15 	&	\nodata	 &	\nodata	 &  &	-3.84 	&	16	&  &	\nodata	 &	\nodata	 &  &	\nodata	 &	\nodata	\\
43.35 	&	\nodata	 &	\nodata	 &  &	-4.75 	&	2	&  &	\nodata	 &	\nodata	 &  &	\nodata	 &	\nodata	\\
\enddata 
\end{deluxetable}
\begin{deluxetable}{cccc}
\tablenum{6}
\tablecolumns{4} 
\tablewidth{0pc} 
\tablecaption{\label{acfdata}Angular two-point correlation function amplitudes for the COSMOS field} 
\tablehead{ 
\colhead{$\theta$} & \multicolumn{3}{c}{$w(\theta)$} \\
\cline{2-4} \\ 
\colhead{(degree)} & \colhead{All} & \colhead{$42.03 \leq \log L < 42.30$} & 
\colhead{$42.30 \leq \log L$} 
}
\startdata
0.0011  &     8.652$\pm$0.255     &   9.755$\pm$0.571    &   14.984$\pm$0.730    \\
0.0018  &     3.035$\pm$0.162     &   2.378$\pm$0.368    &    5.097$\pm$0.459    \\
0.0028  &     1.138$\pm$0.102     &   0.937$\pm$0.229    &    1.762$\pm$0.289    \\
0.0044  &     0.605$\pm$0.064     &   0.726$\pm$0.146    &    1.309$\pm$0.183    \\
0.0070  &     0.335$\pm$0.041     &   0.430$\pm$0.092    &    0.502$\pm$0.117    \\
\enddata 

The complete version of this table is in the electronic edition of
the Journal. The printed edition contains only a sample.
\end{deluxetable}

\end{document}